\documentclass[a4paper,fleqn,usenatbib]{mnras}

\usepackage{newtxtext,newtxmath}

\usepackage[T1]{fontenc}
\usepackage{ae,aecompl}
\usepackage{threeparttable}

\usepackage{graphicx}	
\usepackage{amsmath}	
\usepackage{amssymb}	

\title[Projected alignment of star, gas, and dark matter in galaxy clusters]
{Projected alignment of non-sphericities of stellar, gas, and dark matter
distributions in galaxy clusters: analysis of the Horizon-AGN simulation}

\author[Taizo Okabe et al.]{Taizo Okabe$^{1}$,\thanks{E-mail: taizo.okabe@utap.phys.s.u-tokyo.ac.jp}
Takahiro Nishimichi$^{2}$,
Masamune Oguri$^{1,2,3}$,
S{\'e}bastien Peirani$^{1,4,5}$,
\newauthor
Tetsu Kitayama$^{6}$,
Shin Sasaki$^{7}$, 
and Yasushi Suto$^{1,3}$
\\
$^{1}$Department of Physics, The University of Tokyo, 7-3-1 Hongo, Bunkyo-ku, Tokyo 113-0033, Japan\\
$^{2}$Kavli Institute for the Physics and Mathematics of the Universe (WPI), The University of Tokyo Institutes for Advanced Study, \\
The University of Tokyo, 5-1-5 Kashiwanoha, Kashiwa, Chiba 277-8583, Japan\\
$^{3}$Research Center for the Early Universe, School of Science, The University of Tokyo, 7-3-1 Hongo, Bunkyo-ku, Tokyo, 113-0033, Japan\\
$^{4}$Universit\'e C\^ote d'Azur, Observatoire de la C\^ote d'Azur, CNRS, Laboratoire Lagrange, Bd de l’Observatoire, CS 34229, 06304 Nice Cedex 4, France \\
$^{5}$Institut d'Astrophysique de Paris (UMR 7095: CNRS \& UPMC), 98 bis Bd Arago, 75014 Paris, France \\
$^{6}$Department of Physics, Toho University, Funabashi, 2-2-1 Miyama, Funabashi, Chiba 274-8510, Japan\\
$^{7}$Department of Physics, Tokyo Metropolitan University, 1-1 Minami-Osawa, Hachioji, Tokyo 192-0397, Japan\\
}

\date{Accepted XXX. Received YYY; in original form ZZZ}

\pubyear{2017}

\begin{document}
\label{firstpage}
\pagerange{\pageref{firstpage}--\pageref{lastpage}}
\maketitle

\begin{abstract}
While various observations measured ellipticities of galaxy clusters and alignments between orientations of the brightest cluster galaxies and their host clusters,
there are only a handful of numerical simulations that implement realistic baryon physics to allow direct comparisons with those observations.
Here we investigate ellipticities of galaxy clusters and alignments between various components of them and the central galaxies
in the state-of-the-art cosmological hydrodynamical simulation Horizon-AGN,
which contains dark matter, stellar, and gas components in a large simulation box of $(100 h^{-1}$\,Mpc$)^3$ with high spatial resolution ($\sim1$\,kpc).
We estimate ellipticities of total matter, dark matter, stellar, gas surface mass density distributions, X-ray surface brightness, and the Compton $y$-parameter of the Sunyaev-Zel'dovich effect,
as well as alignments between these components and the central galaxies for 120 projected images of galaxy clusters with masses $M_{200}>5\times10^{13}M_{\odot}$.
Our results indicate that the distributions of these components are well aligned with the major-axes of the central galaxies,
with the root mean square value of differences of their position angles of $\sim 20^\circ$,
which vary little from inner to the outer regions.
We also estimate alignments of these various components with total matter distributions,
and find tighter alignments than those for central galaxies with the root mean square value of $\sim 15^\circ$.
We compare our results with previous observations of ellipticities and position angle alignments and find reasonable agreements.
The comprehensive analysis presented in this paper provides useful prior information for analyzing stacked lensing signals
as well as designing future observations to study ellipticities and alignments of galaxy clusters.
\end{abstract}




\section{Introduction}
Galaxy clusters have played a crucial role in establishing the standard cosmological model and in constraining cosmological parameters.
For example, various observable of galaxy clusters are often used to constrain cosmological parameters 
such as the number density of galaxy clusters in 
X-ray \cite[e.g.][]{2009ApJ...692.1060V, 2014A&A...570A..31B, 2014MNRAS.440.2077M, 2016ApJ...832...95D, 2017MNRAS.469.3738S},
optical \cite[e.g.][]{2010ApJ...708..645R, 2014MNRAS.439.1628Z, 2015PASJ...67...34H, 2015MNRAS.450.2888L, 2018MNRAS.474.1116S},
radio \cite[e.g.][]{2013ApJ...763..127R, 2013ApJ...763..147B, 2014A&A...571A..20P, 2016A&A...594A..24P, 2017MNRAS.469..394H},
baryon fraction in clusters \citep[e.g.][]{2004MNRAS.353..457A, 2006ApJ...652..917L, 2008MNRAS.383..879A, 2011Sci...331.1576S},
and joint analysis of diameter distances for X-ray surface brightness and the Sunyaev-Zel'dovich effect \citep[e.g.][]{2004MNRAS.352.1413S, 2006ApJ...647...25B, 2015MNRAS.447..479W}.
In such cosmological studies, the sphericity of dark matter haloes of galaxy clusters is usually assumed.

However, observations suggest that the shape of galaxy clusters is more like triaxial rather than spherical  
which has been measured by using various estimators such as
the distribution of member galaxies
\citep[e.g.][]{1993ApJ...416..546F, 2005MNRAS.359..191S, 2009ApJ...700.1686P, 2015ApJ...813...20B, 2017arXiv170511167S},
weak lensing 
\citep[e.g.][]{2009ApJ...695.1446E, 2010MNRAS.405.2215O, 2012MNRAS.420.3213O, 2016MNRAS.457.4135C, 2017MNRAS.467.4131V},
strong lensing 
\citep[e.g.][]{2010MNRAS.404..325R},
X-ray surface brightness
\citep[e.g.][]{2010ApJ...719.1926K, 2012ApJ...755..116L, 2015A&A...575A.127P, 2017ApJ...846...51L}
or Sunyaev-Zel'dovich (SZ) effect 
\citep[e.g.][]{donahue}.
Furthermore, many numerical simulations have also reported that the shape of dark matter haloes is approximately triaxial \citep[e.g.][]{2015MNRAS.454.3341C, 2016MNRAS.456.2486D, 2017MNRAS.466..181D, 2017MNRAS.469.1997D, 2017MNRAS.467.3226V}, and their axis ratios depend on cosmological parameters 
\citep[e.g.][]{2003ApJ...588....7S, 2004astro.ph..5097R, 2006ApJ...647....8H}.
Therefore, the triaxiality or ellipticity of galaxy clusters should be taken into account to estimate cosmological parameters more accurately from galaxy clusters.

In addition, simulations suggest that the ellipticity of galaxy clusters can be used to test various physics.
For example, the shape of dark matter distributions in galaxy clusters are affected by implemented baryon physics \citep[e.g.][]{2010MNRAS.404.1137B, 2015MNRAS.452..343S, 2013MNRAS.429.3316B, suto17}, which would be much rounder in all scales up to $\sim1$\,Mpc without the feedback effect from the active galactic nucleus (AGN) \citep[e.g.][]{suto17}.
Self interacting dark matter models predict more spherical distributions in the inner region of galaxy clusters 
than collisionless dark matter \citep[e.g.][]{2000ApJ...535L.103Y, 2000PhRvL..84.3760S, 2001ApJ...547..574D, 2010ARA&A..48..495F, 2013MNRAS.430...81R, 2013MNRAS.430..105P, 2017arXiv170502358T}.
Modified gravity theories generally predict more spherical mass distributions at scales larger than member galaxy distributions \citep[e.g.][]{2013JCAP...10..012H, 2015PhRvD..91b4022K, 2017MNRAS.468.3174L}.

The alignment of major axes of matter distributions in galaxy clusters is also useful for testing
the structure formation scenario in standard $\Lambda$-dominated cold dark matter ($\Lambda$CDM) model.
The $\Lambda$CDM model predicts the hierarchically structure formation, which results in the existence of coherent structures well aligned at various scales.
Observationally, the alignment between major axes of central galaxies and those of member galaxy distributions in galaxy clusters was first recognized by \cite{1968PASP...80..252S},
and measured in detail by \cite{1982A&A...107..338B}.
Many observations have reported the alignment 
between central galaxies and distributions of member galaxies 
\citep[e.g.][]{1986AJ.....91..471A, 1987nngp.proc..227D, 1987A&A...183..217R, 1988AJ.....95..996L, 2017NatAs...1E.157W},
central galaxies and X-ray surface brightness (XSB) distributions 
\citep[e.g.][]{1991A&A...243...38R, 1991AJ....101.1561P, 2008MNRAS.390.1562H}, 
central galaxies and weak lensing signals 
\citep[e.g.][]{2016MNRAS.457.4135C, 2017arXiv170511167S, 2017MNRAS.467.4131V},
distributions of member galaxies and weak lensing signals 
\citep[e.g.][]{2009ApJ...695.1446E, 2017MNRAS.467.4131V},
strong lensing and weak lensing \citep{2012MNRAS.420.3213O},
and among XSB, SZ, lensing signals, and central galaxies \citep{donahue}.

Numerical studies based on $N$-body simulation also suggest the existence of the alignment between central sub-haloes and host clusters 
\citep[e.g.][]{1998ApJ...502..141D, 2008ApJ...675..146F, 2012JCAP...05..030S, 2012ApJ...748...98S}.
However, there are only a few numerical studies that estimate the alignment between central galaxies or gas distributions and dark matter haloes of host clusters \citep[e.g.][]{2015MNRAS.453..721V, 2015MNRAS.453..469T}, given that dark matter only $N$-body simulations cannot derive position angles of central galaxies.
While simulations with baryon effects are needed to investigate such alignment,
they are challenging because baryon processes such as gas dynamics, star formation, and feedback are too complicated to be fully implemented.
In addition, to analyse the alignment statistically we need both high resolution to resolve central galaxies and a large simulation box to contain sufficient number of galaxy clusters.

The Horizon-AGN simulation \citep{dubois14} is a state-of-the-art hydrodynamical cosmological simulation that enables us to investigate the alignment at high precision.
It simulation solves the evolution of dark matter, stellar, and gas components in a large simulation box of $(100 h^{-1}$\,Mpc$)^3$ with high spatial resolution ($\sim1$\,kpc).
It simulation contains many galaxies and galaxy clusters and can explain various observations from galaxy ($\sim10$\,kpc) to galaxy cluster scales ($\sim1$\,Mpc) \citep[e.g.][]{2016MNRAS.463.3948D, 2017MNRAS.472.2153P}.
There are previous studies to estimate the ellipticities and alignments of different components using the Horizon-AGN simulation.
\cite{suto17} measured ellipticities of projected distributions but they did not study the alignments among different components. 
\cite{2017MNRAS.472.1163C} investigated the alignment between galaxies and their host dark matter haloes.
In this paper, we present a more comprehensive study of projected ellipticities and position angles of various components, including the central galaxies, dark matter, stellar, gas, total surface mass density, XSB, and SZ proxies of galaxy clusters.
Our theoretical predictions based on the realistic hydrodynamical simulation should provide useful guidance for interpreting various observations of ellipticities and alignments as well as designing future observations on this topic.

The structure of this paper is as follows.
We summarize the Horizon-AGN simulation in Section~\ref{sec:simulation}, and describe our fitting procedure in Section~\ref{sec:ell}.
In Section~\ref{sec:corr}, we discuss correlation of ellipticities and position angles among different components.
We present statistical values of ellipticities and position angles in Section~\ref{sec:stat},
and compare them with observations in Section~\ref{sec:obs}.
We summarize our results in Section~\ref{sec:summary}.
In Section Appendix \ref{sec:app}, we show representative images of three clusters to show the morphological diversity of galaxy cluster that 
we analysed in this paper.

\section{Identifying galaxies and clusters in the Horizon-AGN simulation} \label{sec:simulation}
The Horizon simulations consist of three simulations, Horizon-AGN, Horizon-noAGN, and Horizon-DM. 
In this paper, we use the Horizon-AGN simulation, although there is another cosmological hydrodynamical simulation, Horizon-noAGN (see \citealt{2017MNRAS.472.2153P}).
The Horizon-noAGN simulation adopts exactly the same initial condition and physical process 
except for AGN feedback.
\cite{suto17} compared axis ratios of galaxy clusters in the Horizon-AGN with those in Horizon-noAGN
to find large effects of AGN feedback on the axis ratios.
However, \cite{suto17} also found that AGN feedback is important to match simulations with 
various observations such as mass density profiles, temperature profiles, and ellipticities of galaxy clusters.
This is why we focus on the Horizon-AGN simulation in the paper.

\subsection{Horizon-AGN Simulation} \label{sec:horizon}

We examine the correlations of non-sphericities of projected surface densities among different components of simulated galaxy
clusters. In particular, we are interested in the alignment of their position angles with respect to those of central galaxies. 
Clearly this requires a cosmological hydrodynamical simulation implemented with detailed baryon physics 
and also with high spatial and mass resolutions to identify central galaxies in the cluster centres.  
We thus focus on the Horizon-AGN simulation, the state-of-the-art cosmological hydrodynamical simulation.
The detail of this simulation is already described in \cite{dubois14}. 
Thus we summarize only its major features relevant to our current work.

The Horizon-AGN simulation adopts the standard $\Lambda$CDM cosmological model.
The cosmological parameters are based on the seven-year Wilkinson
Microwave Anisotropy Probe \citep{WMAP}; $\Omega_{{\rm m}, 0} = 0.272$ (total matter density at present day), 
$\Omega_{\Lambda, 0} = 0.728$ (dark energy density at present day), 
$\Omega_{{\rm b}, 0} = 0.045$ (baryon density at present day), 
$\sigma_{8} = 0.81$ (amplitude of the power spectrum of density fluctuations that are averaged on spheres of $8h^{-1}$ \,Mpc radius at present day), 
$\rm H_{0} = 70.4$\,km/s/Mpc (Hubble constant), and $n_{s} = 0.967$
(the power-law index of the primordial power spectrum).

The simulation is performed in a periodic cube of $(100 h^{-1}$\,Mpc$)^3$, and the initial condition is generated with MPGRAFIC software \citep{2008ApJS..178..179P}.  
The simulation follows the evolution of three different components, dark matter, gas, and star.
Dark matter is represented by $N=1024^{3}$ equal-mass particles in the entire box, corresponding to the mass resolution of $8.27\times10^{7}$\,M$_{\odot}$.  
Baryon gas is assigned over the meshes in the simulation box, and its evolution is solved with the adaptive mesh refinement code RAMSES \citep{2002A&A...385..337T}.  
Star is represented by collisionless particles, whose formation is modeled on the basis of an empirical Schmidt law.
Since those star particles are created according to a random Poisson process, their masses are not the same, but typically around $2\times 10^6 M_\odot$.

The evolution of collisionless particles (dark matter and star) are followed by the particle-mesh solver with a cloud-in-cell interpolation. 
Therefore the spatial resolution depends on the size of the local cell where those particles are located.  
The initial size of the gas cell is $136$\,kpc, and then refined up to 1.06\,kpc ($=136/2^7$kpc after seven times refinement), 
which corresponds to the highest spatial resolution achieved in the simulation.

In addition to radiative cooling and hydrodynamical evolution of
gas component, feedback from stars is implemented assuming the Salpeter initial mass
function \citep{1955ApJ...121..161S} with lower and upper mass limits of $0.1M_\odot$ and $100M_\odot$, respectively.  
The mechanical energy from Type II supernova explosions and stellar winds is computed according to the STARBURST99
\citep[]{1999ApJS..123....3L, 2010ApJS..189..309L} with the frequency of Type Ia supernova explosions computed using \cite{1983A&A...118..217G}.

There are two different AGN feedback modes in the Horizon-AGN simulation; one is {\it radio} mode and the other is {\it quasar} mode depending on the Eddington ratio
$\chi \equiv \dot{M}_{\rm BH}/\dot{M}_{\rm Edd}$, where $\dot{M}_{\rm BH}$ is the accretion rate onto black holes 
and $\dot{M}_{\rm Edd}$ is the effective upper limit of the accretion (Eddington accretion rate).
Recent observations \citep[e.g.][]{2016Natur.533..504C} support that both modes exist.
At low accretion rate $\chi < 0.01$, feedback from black holes behaves as radio mode
which injects the energy into a bipolar outflow with a jet velocity of $10^4$\,km s$^{-1}$.
The outflow jet is modeled as a cylinder following \cite{2004MNRAS.348.1105O}.
\cite{2010MNRAS.409..985D} describes more details.
The energy deposition rate of the radio mode is computed by $\dot{E}_{\rm AGN} = \epsilon_{\rm f}\epsilon_{\rm r}\dot{M}_{\rm BH}c^2$ 
where $\epsilon_{\rm f}$ is the free parameter, $\epsilon_{\rm r}$ is the radiative efficiency, and $c$ is the speed of light.
The radiative efficiency is assumed to be equal to $\epsilon_{\rm r} = 0.1$ following \cite{1973A&A....24..337S}, and $\epsilon_{\rm f}$ is set to unity for the radio mode.
The quasar mode is adopted at high accretion rate $\chi > 0.01$, which deposits the thermal energy into the gas isotropically
at an energy deposition rate $\dot{E}_{\rm AGN}$.
The free parameter $\epsilon_{\rm f}$ is chosen so as to reproduce various observations such as 
the scaling relations between black hole masses and galaxy properties 
(bulge masses and velocity dispersions of stars) and the black hole density in our local Universe.
The details are given in \cite{2012MNRAS.420.2662D}.

The dataset from the Horizon-AGN simulation has been examined in detail by various authors from different aspects, and has been shown
to reproduce well observed properties such as intrinsic alignment of galaxies \citep{2015MNRAS.454.2736C, 2016MNRAS.461.2702C},
morphological diversity of galaxies, galaxy-halo mass relation, size-mass relation of galaxies \citep{2016MNRAS.463.3948D},
AGN luminosity function, black hole mass density \citep{2016MNRAS.460.2979V}, density profile of massive galaxies \citep{2017MNRAS.472.2153P, 2018arXiv180109754P},
high-mass end of the galaxy stellar mass function \citep{2017MNRAS.472..949B}, 
luminosity functions of galaxies, stellar mass functions, the star formation main sequence, rest-frame UV-optical-NIR colours,
the cosmic star formation history in the redshift range $1<z<6$ \citep{2017MNRAS.467.4739K}, ellipticities of X-ray galaxy clusters \citep{suto17},
and tight relation between black hole masses in the brightest group/cluster galaxies and their host group/cluster masses \citep{2017arXiv171109900B}.
These properties are not accounted for in the Horizon-noAGN.

\subsection{Locating Galaxy Clusters and Central Galaxies} \label{sec:cluster} 
We identify dark matter haloes using the ADAPTAHOP halo finder \citep[]{2004MNRAS.352..376A, 2009A&A...506..647T}.
We pick up all cluster sized dark matter haloes with masses larger than $5\times10^{13}M_{\odot}$.
The masses of dark matter haloes are defined by those within spherical average density 
larger than 200 times the critical density of the Universe.
We regard these haloes as galaxy clusters.
The total number of these galaxy clusters is 40 in the Horizon-AGN simulation.

The definition of the centre of each cluster needs to be considered carefully as well. 
One reasonable option is to compute the centre-of-mass for each cluster from their dark matter, star and gas
components. This is a straightforward procedure in simulation data, but is difficult to apply in observations.
In reality, the centre of observed galaxy clusters is often defined as
the location of its brightest cluster galaxy (BCG).  
While we can compute the luminosity of each galaxy in principle \citep{dubois14}, it is complicated and also subject to uncertainty of the star formation history.
Therefore we decide to adopt the position of the most {\it massive} galaxy 
(within 1 Mpc from the most bound particle of each halo) as the centre of cluster, 
where galaxies are identified with the ADAPTAHOP finder applying to stellar particles. 
In this paper, we call such a galaxy as central galaxy (CG) so as to distinguish it from BCG. 
In practice, however, they are supposed to be almost identical to the observed BCGs. 
Thus we identify these two populations when we compare our results with the observation in Section~\ref{sec:obs}.

Since the observational data provide only projected images,
we focus on the projected alignment between CG and other components in our simulation.
First we determine the position angle and ellipticity of projected CG. 
Following \cite{suto17}, we use the mass tensor to estimate the ellipticity and the position angle:
\begin{equation}
    I_{{\rm CG}, \alpha\beta} \equiv 
     \sum^{N_{\rm CG}}_{i=1} m_{\rm CG}^{(i)}
     \left[x^{(i)}_{{\rm CG}, \alpha} - x_{{\rm CG}, \alpha}^{\rm CM}\right]
     \left[x^{(i)}_{{\rm CG}, \beta}   - x_{{\rm CG}, \beta}^{\rm CM}\right]
     {~~~~~~~}(\alpha, \beta = 1,2),
\label{eq:bcgtensor}
\end{equation}
where $m_{\rm CG}^{(i)}$ and $x^{(i)}_{{\rm CG}, \alpha} - x_{{\rm CG}, \alpha}^{\rm CM}$ are the mass and the projected
position vector of the $i$-th CG particle relative to the centre of mass, respectively.
The summation runs over the $N_{\rm CG}$ star particles within the ellipse whose
size is $\sqrt{ab} = 20$\,kpc, where $a$ and $b$ are the semi-major and semi-minor axes, respectively.
We diagonalize the mass tensor to obtain semi-major axis $a$, semi-minor axis $b$, and position angle $\theta$.
We start from a circle with radius of $r=20$\,kpc centred at the centre of mass of CG particles.
Then we reset the centre of mass of particles within the new ellipse
and compute the tensor iteratively until both eigenvalues of the tensor are converged within $10^{-8}$.
The bottom panels in Fig.~\ref{fig:im2}, \ref{fig:im3}, and \ref{fig:im7} show examples of resulting ellipses for three galaxy clusters. 

When both eigenvalues converge, we obtain values of semi-major axis $a_{\rm CG}$, semi-minor axis $b_{\rm CG}$, 
the centre of mass $x_{{\rm CG}, \alpha}^{\rm CM}$,  and position angle $\theta_{\rm CG}$ of the CG.
We define the ellipticity of the CG as:
\begin{equation}
\epsilon_{\rm CG}=1-\frac{b_{\rm CG}}{a_{\rm CG}}.
\end{equation}

For each cluster, we consider three different projection directions assuming $x$-, $y$-, and $z$-axes as line-of-sight directions.
We regard these three projections as independent so that we effectively have $N_{\rm cl}\equiv120$ galaxy clusters for our analysis.
Although the three different projection directions are not independent,
we confirmed that our results such as mean ellipticities  and the rms of position angle differences between various 
components shown in Section~\ref{sec:statradial} are not significantly changed even if we do not combine results with these three different projection directions.

\section{Ellipticity and position angle from projected images of the clusters} \label{sec:ell}
In the Horizon-AGN simulation, dark matter and star are defined by particles but gas is computed in the adaptive mesh.
Each dark matter particle has the same mass and has the position,
whereas each star particle has both a position and mass.
Each adaptive mesh contains position, mass, metallicity, temperature, and size of the mesh.

In this paper, we compare the ellipticity and position angle 
for projected images of X-ray surface brightness (XSB), Compton $y-$parameter (SZ), total surface mass
density (tot), dark matter surface mass density (DM), star surface mass
density (star), and gas surface mass density (gas) of galaxy clusters in the Horizon-AGN simulation.
To create these projected images, we first define a cube with a size of $(4.24\,{\rm Mpc})^3$ centred at the CG.
Note that the position of the CG is defined as the center of mass $x_{{\rm CG}, \alpha}^{\rm CM}$ computed in Section \ref{sec:cluster}. 
Then, we divide the cube into $(4001)^3$ meshes with a size of $(1.06\,{\rm kpc})^3$,
which corresponds to the minimum size of the adaptive mesh in the Horizon-AGN simulation.
Mass densities of dark matter, star and gas, metallicity, and temperature, are assigned to each mesh.
For dark matter and star, mass densities are simply computed by the nearest grid point method,
in which mass of each particle is assigned to the nearest mesh in a projected plane.
Since the gas property is computed in the adaptive mesh, we divide all meshes into the smallest
meshes of $(\Delta = 1.06\,{\rm kpc})^3$ with the same values of temperature, metallicity, and mass density. 
For these projected images, ellipticity and position angle are estimated by using a tensor weighted by projected values
such as surface mass density, XSB, and $y$-parameter as described in the following subsections.

\subsection{Surface densities of different components} \label{sec:image}
Projected images are created as follows:
\begin{enumerate}
\renewcommand{\labelenumi}{(\Roman{enumi})}
\item surface mass density (DM, star, gas, and tot): \\
We compute the mass density of the mesh
\begin{equation}
\rho_{A}(i, j, k) = m_{A}(i, j, k) / \Delta^3,
\label{eq:rho}
\end{equation}
where $0 \leq i, j, k \leq 4000$ are indices 
specifying the mesh, and $m_{A}(i, j, k)$ and $\rho_{A}(i, j, k)$ 
are mass and mass density of $A$
component ($A = $ DM or star or gas) in $(i, j, k)$ mesh, respectively.
The surface mass density is calculated by integrating the mass density along the line of sight:
\begin{equation}
    \Sigma_{A} (i, j) =  \Delta \sum_{k=0}^{4000} \rho_{A}(i, j, k).
\label{eq:smdp}
\end{equation}

The total mass density is simply computed by the summation of all these components,
\begin{equation}
    \Sigma_{\rm tot} (i, j) = 
	  \Sigma_{\rm DM} (i, j) + \Sigma_{\rm star} (i, j) + \Sigma_{\rm gas} (i, j). 
	\label{eq:smdtot}
\end{equation}

\item X-ray Surface Brightness (XSB): \\
The X-ray Surface Brightness (XSB) is calculated as 
\begin{equation}
    \Sigma_{\rm XSB} (i, j) \propto  \sum_{k=0}^{4000} 
    {~} n_{\rm gas}^2(i, j, k) \Lambda(T, Z),
	\label{eq:xsbsim}
\end{equation} 
where $n_{\rm gas}(i, j, k)$, $\Lambda(T, Z)$, $T=T(i, j, k)$, and $Z=Z(i, j, k)$ denote the number density, cooling function, temperature, and metallicity of the gas  
in a mesh specified by $(i, j, k)$, respectively.  
We use the package SPEX \citep{1996uxsa.conf..411K} to derive the cooling function, $\Lambda$,
for the photon energy band, $0.5$\,keV$< E < 10$\,keV.

The molecular number density of the gas is computed from the mass density:
\begin{equation}
    n_{\rm gas}(i, j, k) = \frac{\rho_{\rm gas} (i, j, k)}{\mu m_{\rm p}}
	\label{eq:m-n}
\end{equation}
where $\mu$ and $m_{\rm p}$ represent the mean molecular weight and mass of proton, respectively.
We confirmed the mean molecular weight is almost constant independent of the position of meshes within the range of our interest.
Since we are interested in only the shape of each component, the normalization does not affect our results
and exact value of $\mu$ is not important.

\item Compton $y$-parameter of the Sunyaev-Zel'dovich effect (SZ) : \\ 
The thermal Sunyaev-Zel'dovich effect is characterized by the Compton $y$-parameter.
We calculate the $y$-parameter in the Horizon simulation as follows:
 \begin{equation}
    \Sigma_{\rm SZ} (i, j) \propto \sum_{k=0}^{4000}  n_{\rm gas}(i, j, k) T(i, j, k).
\label{eq:szsim}
\end{equation}
\end{enumerate}

Fig.~\ref{fig:2z} plots an example of the images projected to the $z$-direction for one cluster in our sample.
The further detail of this cluster is described in Section~\ref{sec:no2}.

\subsection{Procedure of ellipse fit} \label{sec:fit}
In order to estimate the ellipticity of each component described in
Section~\ref{sec:image}, we use surface density weighted tensor:
\begin{equation}
I_{A, \alpha\beta}  = \sum_{i, j} \Sigma_{A} (i, j) \left[x_{\alpha} (i, j) - x_{{\rm CG}, \alpha}^{\rm CM}\right]
                                                                        \left[x_{\beta} (i, j)  - x_{{\rm CG}, \beta}^{\rm CM} \right]
\label{eq:tensor}
\end{equation}
where $x_{\alpha} (i, j) - x_{{\rm CG}, \alpha}^{\rm CM}$ and $\Sigma_{A} (i, j)$ denote the projected position relative to centre of mass and value of $(i, j)$ cell, respectively.
The summation runs over cells within a given enclosed ellipse region.

We basically follow \cite{2016PASJ...68...97S, suto17} to estimate ellipticities and position angles.
However, we fix the centre of the ellipse to that of CG, $x_{{\rm CG}, \alpha}^{\rm CM}$
derived in Section~\ref{sec:cluster}, unlike those papers where they set the centre to the centre of mass
since we are especially interested in the ellipticity and position angle that can be directly compared with observations.

We diagonalize the tensor to obtain values of axis ratio $b/a (<1)$ and position angle.
We define the ellipticity as $\epsilon \equiv 1- b/a$.
Starting from a circle with radius $r$, the above process is iterated changing the axis ratio $b/a$
until both two eigenvalues of the tensor converge within $10^{-8}$.
We confirm that both values of ellipticity and position angle converge well by this convergence criteria.
When both eigenvalues converge, we obtain the final values of ellipticities $\epsilon_{A}$, and position angles $\theta_{A}$.
Finally, we repeat the same analysis for each galaxy cluster with different sizes of the ellipse, $\sqrt{ab}=0.1$, $0.2$, $\cdots$, $1.0$ \,Mpc.

\subsection{An example of the ellipse fit} \label{sec:no2}
\begin{figure*}
	\includegraphics[width=2\columnwidth]{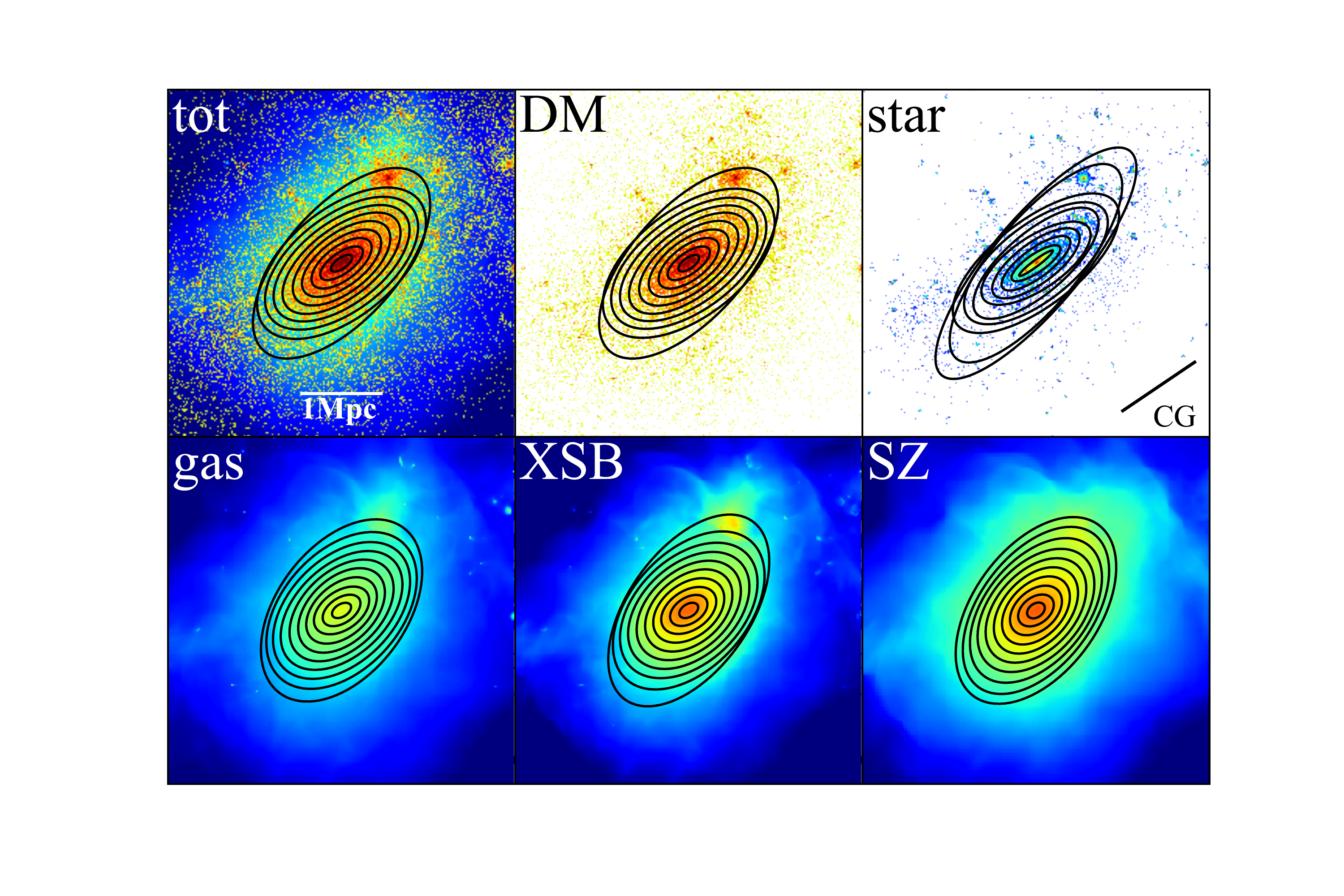}
\caption{An example of projected images of a cluster over 4.24
  Mpc$\times$4.24 Mpc for different components (integrated over 4.24Mpc along the $z$-direction of simulation); 
  total density ({\it upper-left}), dark matter density ({\it upper-centre}), star density ({\it upper-right}), gas density ({\it lower-left}), 
  X-ray surface brightness ({\it lower-centre}), and $y$-parameter from the SZ effect ({\it lower-right}). 
  Those quantities are sampled in $1.06$\,kpc$\times1.06$\,kpc pixels before integrated along the line-of-sight.
  Colour-coded according to their absolute values.  
  Solid curves indicate to ellipses computed by the tensor method described in Section~\ref{sec:fit}, corresponding to $\sqrt{ab}=0.1$, $0.2$, $\cdots$, $1.0$ \,Mpc 
(i.e., the area of each ellipse is $\pi ab$)
The direction of the major-axis of CG is also shown at the lower right in star image.
}
    \label{fig:2z}
\end{figure*}
\begin{figure*}
	\includegraphics[width=2\columnwidth]{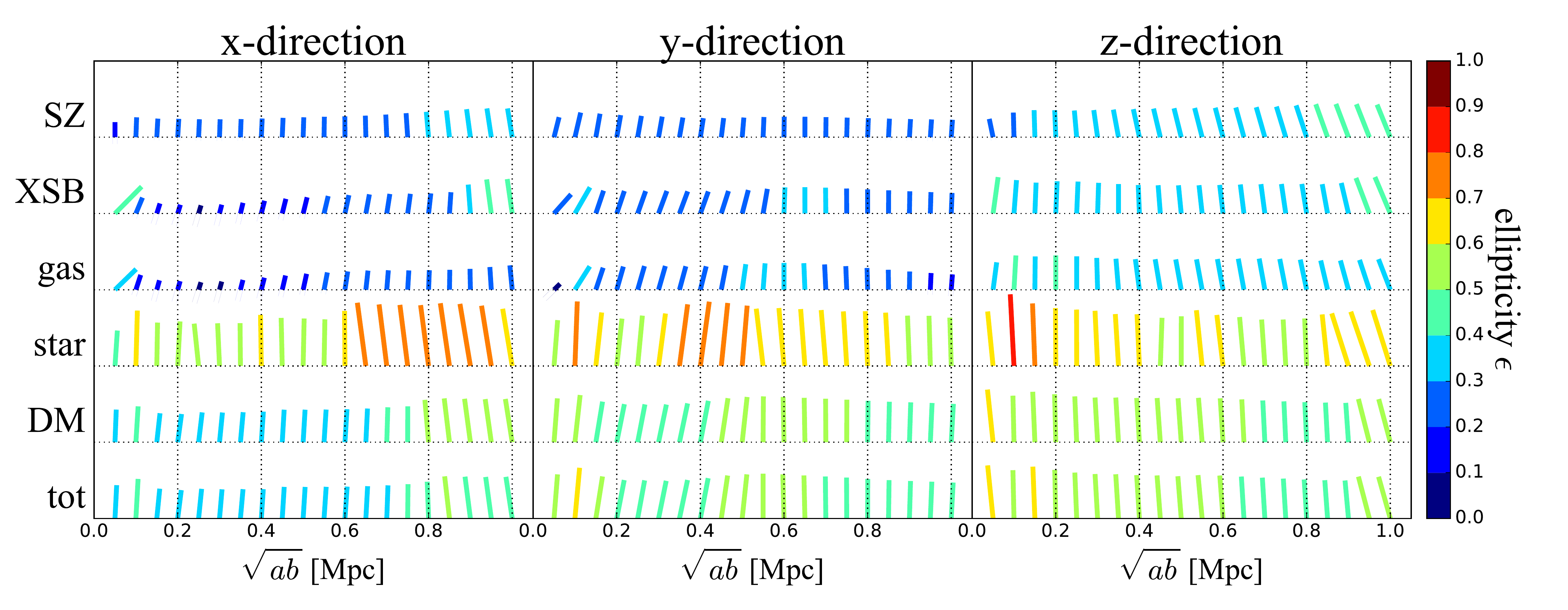}
\caption{Projected position angles of each component of a cluster plotted in Fig.\ref{fig:2z}.  
   The inclination of the bar with respect to the vertical direction indicates the position angle, i.e., the direction of 
   the major axis of the ellipse of each component relative to that of CG.  
  The length and colour of bars denote to the value of the ellipticity $\epsilon = 1-b/a$.  
  Left, middle, and right panels show the result for thee projection along $x$-, $y$-, and
  $z$-directions of the simulation, respectively.}
    \label{fig:arrow}
\end{figure*}
In this subsection, we discuss the resulting images and ellipses derived by the above procedure
for the cluster as an example.
We select the same cluster as illustrated in Fig.~3 of \cite{suto17}, which is the most massive single-core 
dominated cluster with mass of $M_{200} = 6.2\times10^{14}M_{\odot}$.  
\footnote{
Values of $r_{200}$ and $M_{200}$ in \cite{suto17} are incorrectly estimated and 
they are smaller by a factor of 1.5 and 1.4 than true values, respectively.
}
Fig.~\ref{fig:2z} shows images projected along the $z$-direction of the simulation box
for six components (tot, DM, star, gas, XSB, and SZ).
Since we are interested only in the shape of cluster, the absolute values of colour scales are not shown. 
Position angles for the six components at all scales are roughly aligned relative to that of CG.
This is one of our main results, which will be discussed more statistically in Sections~\ref{sec:corr} and \ref{sec:stat}.
Comparing the ellipses for the six components, the stellar density distribution is more elongated,
while those of gas components (gas, XSB, and SZ) are more spherical than that of DM.
The former is because stellar components suffer from strong radiative cooling.
The latter is because the gas distribution follows the gravitational potential of the host cluster
that is rounder than the matter distribution.
Total matter density distribution is almost the same as that of dark matter,
simply because total matter density is dominated by dark matter.

We also evaluate the differences among different projection directions.
Fig.~\ref{fig:arrow} simultaneously plots ellipticity and position angle for each component as a function of scale $\sqrt{ab}$.
The above statements for $z$-direction hold also for the other projections, $x$-, and $y$-directions;
the position angles are clearly aligned with respect to the CG at almost all scales,
gas components are more circular and stellar components are more elongated than that of dark matter, 
and finally the density distribution of total matter is quite similar to that of dark matter. 
The ellipticity does not change substantially against the scales except for that of stellar distribution which is sensitive to the presence of substructures.
Since these results are just derived one cluster, we examine these features more statistically using all the 40 clusters in the next Section.

\section{correlation of ellipticities and position angles among different components} \label{sec:corr}
\subsection{Alignment of position angles} \label{sec:corrpa}
\begin{figure*}
	\includegraphics[width=2.2\columnwidth]{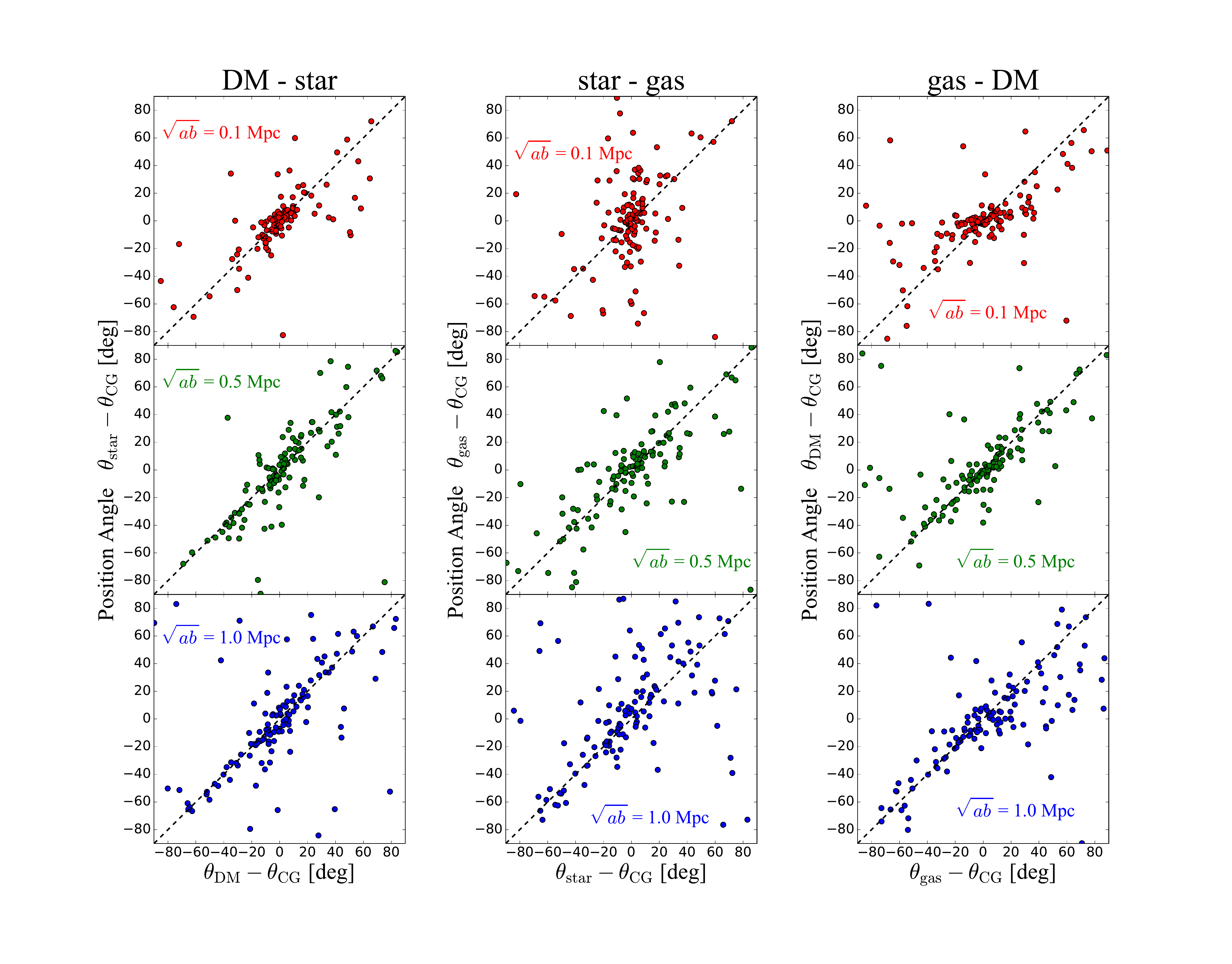}
\caption{Correlation of position angles relative to the CG for different components 
 evaluated at $\sqrt{ab}=0.1$ ({\it top}), $0.5$ ({\it middle}), and $1.0$\,Mpc ({\it bottom}).
Left, middle, and right panels show the correlations between DM and star, star and gas, and gas and DM, respectively. 
}
    \label{fig:pascatter}
\end{figure*}
We pay particular attention to position angles with respect to the CG and among components
to understand the correlation of matter density distributions.
Fig.~\ref{fig:pascatter} plots the correlations of position angles relative to the CG for different components evaluated at $\sqrt{ab}=0.1$, $0.5$, and $1.0$\,Mpc.
If density distributions are aligned with the CG, symbols are expected to be clustered around the origin $(0, 0)$.
For all the three components, the position angles are clustered at the origin
indicating that these density distributions are well aligned with the major-axis of the CG.
At $\sqrt{ab}=0.1$\,Mpc, symbols are more clustered around the origin than at other scales,
which indicates that all the components are relatively well aligned in the inner region.
The distributions of the alignments relative to the CG is consistent with the result for the cluster described in Section~\ref{sec:no2}. 
Incidentally, Fig.~\ref{fig:pascatter} also indicates the alignment of position angles among different components 
even if outer region where the alignments of position angles relative to the CG are worse. We discuss this point more detail below in Section~\ref{sec:statradial}.

\subsection{Correlation of ellipticities} \label{sec:correll}
\begin{figure*}
	\includegraphics[width=1.8\columnwidth]{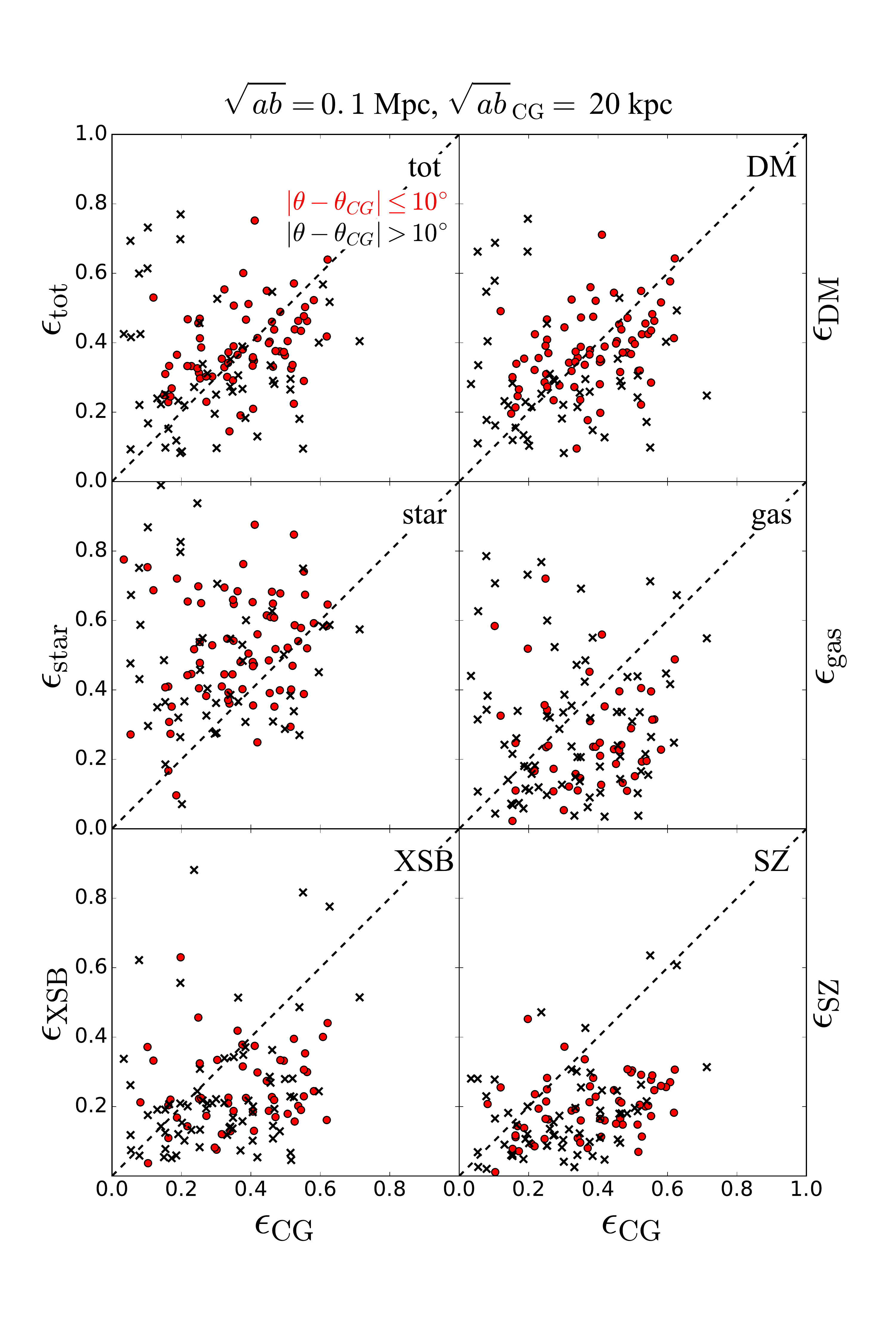}
\caption{Correlation of ellipticities of 
different components evaluated at $\sqrt{ab}=0.1$\,Mpc against that of the CG.
Red and black symbols indicate those with the position angle relative to the CG
of  $\Delta\theta <10^\circ$ and $>10^\circ$, respectively.}
    \label{fig:bcgscatter}
\end{figure*}
Since inferring the density distribution of gas from observational data is generally difficult,
we also consider XSB and SZ, which are directly observable. 
For the similar reason, we also consider total matter density, which can be estimated from lensing analysis.

Fig.~\ref{fig:bcgscatter} shows scatter plots for different components evaluated at $\sqrt{ab}=0.1$\,Mpc with that of the CG.
There are no tight correlations of ellipticities between these components and the CG.
Neither the ellipticities of matter density distribution (DM, star, tot) nor
those of gravitational potential shape (gas, XSB, SZ) correlate with that of the CG. 
This result is inconsistent with a previous work by \cite{2015A&A...581A..31S}, they reported tight correlation between ellipticities of BCG and those of light distributions.
This discrepancy might be due to difference of method used to estimate ellipticities.
They created the light map of galaxy clusters by smoothing light distributions of each member galaxy.
Thus, their ellipses are not affected by each galaxy whereas those derived from our tensor method are affected by each galaxy as illustrated in Fig.~\ref{fig:im7}.

Ellipticities of stellar components are systematically higher than those of the CG.
This is simply due to the other galaxy near the CG.
In fact, an ellipse of $\sqrt{ab}=0.1$\,Mpc (the most inner one) in stellar image of Fig.~\ref{fig:2z} is elongated 
toward a nearby galaxy (bottom left from the CG), which is located along the major-axis of the CG.
Fig.~\ref{fig:pascatter} also indicates that the position angles of stellar component are well aligned with major-axis of the CG 
in spite of no tight correlation of ellipticities between stellar components and the CG.
The alignment suggests member galaxies are preferentially distributed along major-axis of the CG,
which is consistent with previous findings \citep[e.g.][]{1994MNRAS.268...79W, 1995ApJ...451L...5W, 2000ApJ...543L..27W, 2005ApJ...628L.101B, 2006MNRAS.369.1293Y, 2007MNRAS.378.1531K, 2007MNRAS.376L..43A, 2007ApJ...662L..71F, 2013ApJ...768...20L, 2015MNRAS.454.3328V, 2016MNRAS.463..222H, 2017MNRAS.466.4875L, 2017A&A...601A.145F}.

Ellipticities of dark matter and total matter distributions are located around the diagonal line despite with large scatters.
The correlations might be affected by two dominant effects;
one is the projection effect, and the other is the effect of substructure.
The projection effect is explained as follows.
While dark matter and total matter distributions are projected by a length of $4.24$\,Mpc,
ellipticities of CG are computed by using only CG particles that extend only $\sim100$\,kpc along the line-of-sight.
The projections of such a wide length scale for dark matter and total matter make their shapes of surface densities 
rounder than those projected only inner part.
On the other hand, the existence of substructures, which are located preferentially along the major-axis of the CG, 
enhances ellipticities as discussed above.
As a result of these two competitive effects, 
ellipticities of dark matter and total matter distributions may be comparable with those of the CG.

\begin{figure*}
	\includegraphics[width=2.3\columnwidth]{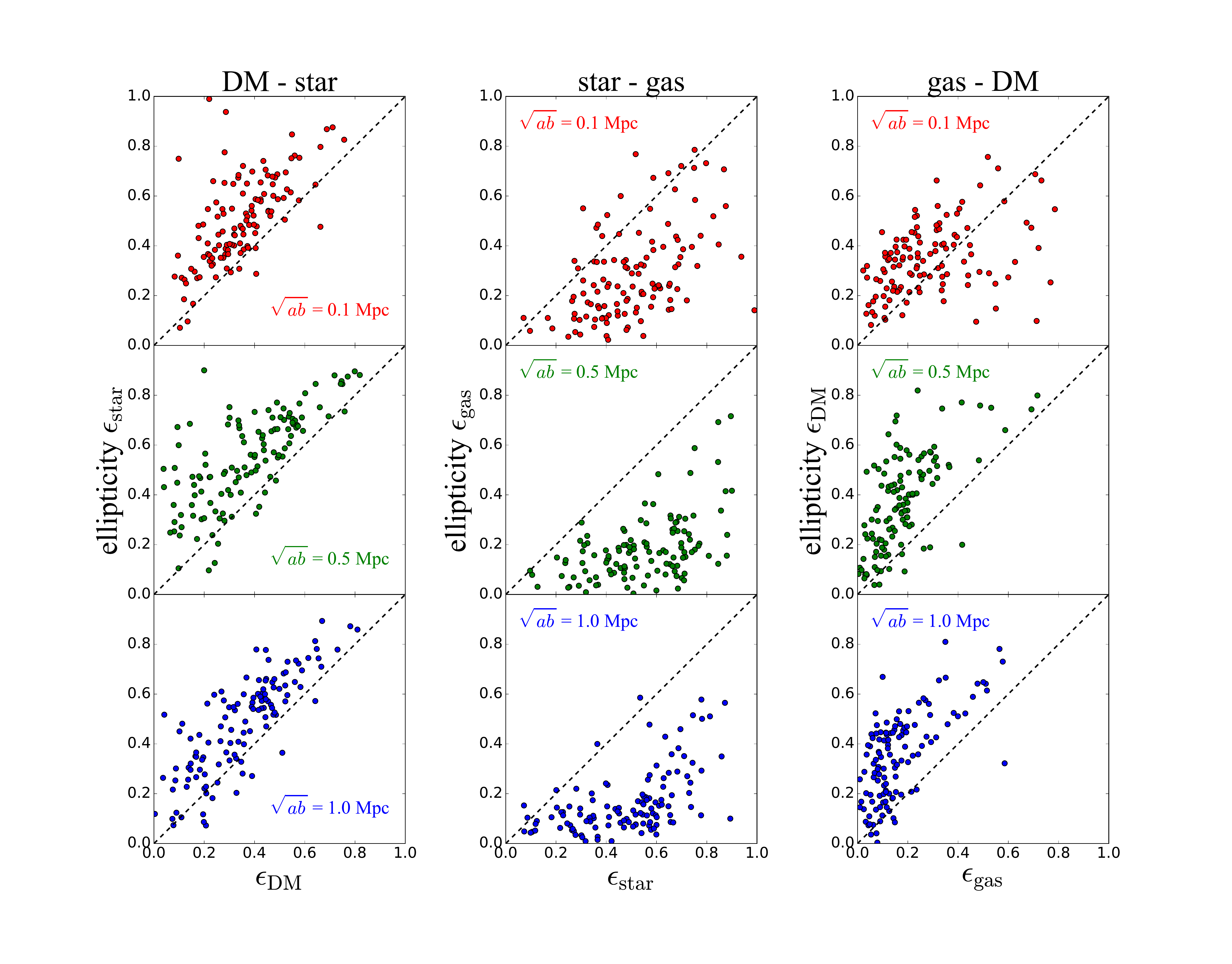}
\caption{
Correlations among ellipticities of different components evaluated 
at $\sqrt{ab}=0.1$ ({\it top}), $0.5$ ({\it middle}), and $1.0$\,Mpc ({\it bottom}).
Left, middle, and right panels show the correlations between DM and star, star and gas, and gas and DM, respectively. 
}
    \label{fig:scatterell}
\end{figure*}
Fig.~\ref{fig:scatterell} plots the correlations among ellipticities of different components 
evaluated at $\sqrt{ab}=0.1$, $0.5$, and $1.0$\,Mpc.
We find that ellipticities of stellar density distributions are higher, 
and those of gas are lower than those of dark matter.
This result is consistent with that for the cluster explained in Section~\ref{sec:no2}.

The strong correlation between DM and star is simply because each dark matter substructure 
contains stellar components that correspond to member galaxies in observations.
In fact, Fig.~\ref{fig:2z} indicates that there is a substructure both in DM and star 
at upper right from the centre, and the ellipse is elongated toward the substructure.

\section{Statistics of cluster shape} \label{sec:stat}
\subsection{Histograms of ellipticity and position angle} \label{sec:pdf}
\begin{figure*}
	\includegraphics[width=1.8\columnwidth]{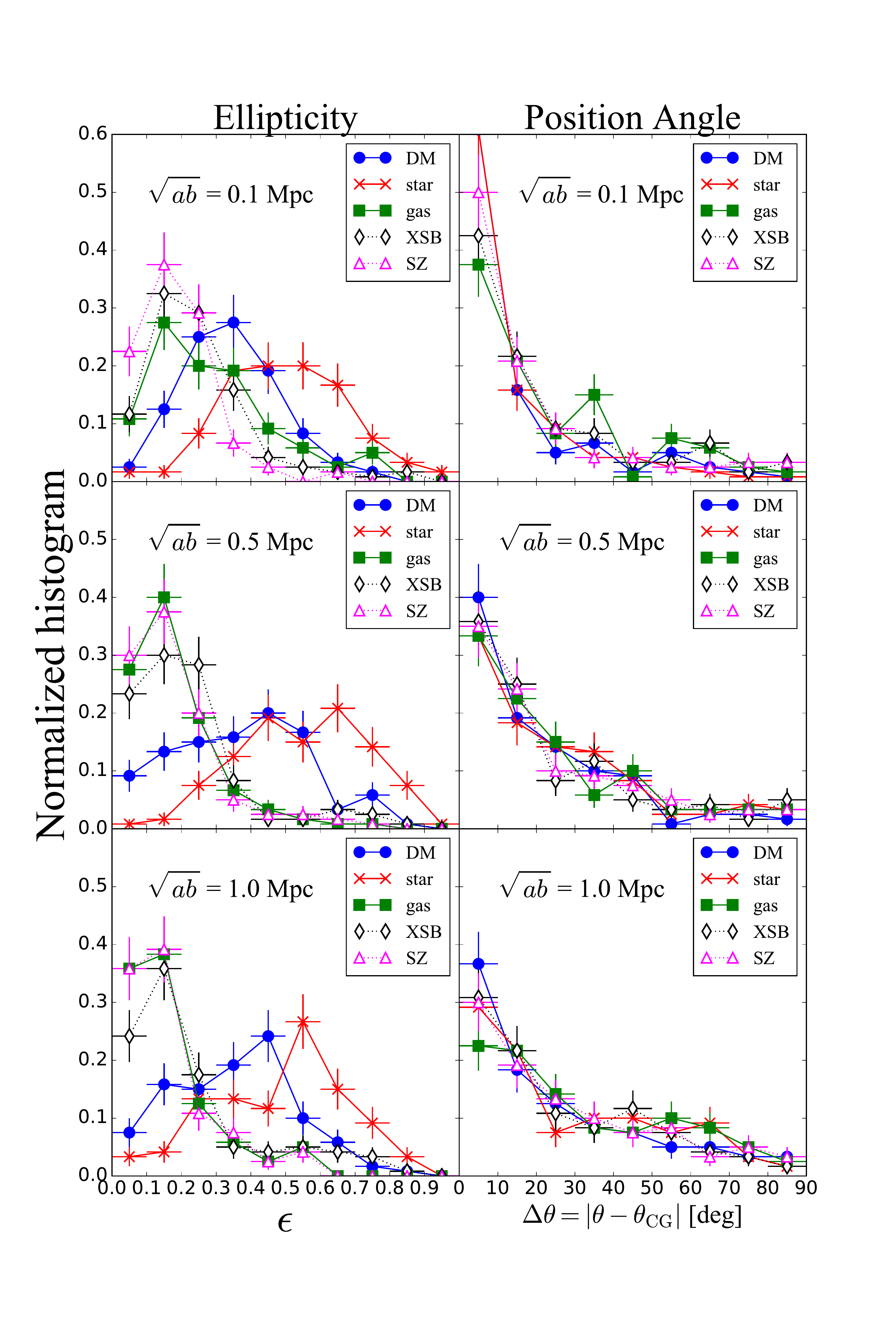}
\caption{Normalized binned distribution functions of ellipticities ({\it Left}) and position angles relative to the CG ({\it Right}) for different components,
which is computed by dividing the number of clusters in each bin by the total number of clusters $N_{\rm cl} =120$.
  Top-, middle-, and bottom panels indicate the results evaluated at $\sqrt{ab}=0.1$, $0.5$, and $1.0$ Mpc of the fitted ellipses, respectively. 
  The vertical and horizontal bars associated with
  each symbol indicate the size of bin and the square root of the number of clusters in each bin (40 different clusters projected along 3 directions).  }
    \label{fig:pdf}
\end{figure*}
Fig.~\ref{fig:pdf} shows normalized histograms of the ellipticities and position angles relative to the CG.
Note that these histograms are computed from $N_{\rm cl}=120$ clusters (40 different clusters projected along three directions).
Clearly the mean value of ellipticity of stellar (gas) distribution is higher (lower) than that of dark matter at all scales.
Histograms of ellipticities for XSB and SZ are quite similar to that of gas.
This result is consistent with that of \cite{suto17},
although the direct comparison is difficult because of slightly different method used for ellipse fitting.
The histograms of position angles are peaked at $\Delta\theta\equiv|\theta-\theta_{\rm CG}|=0$,
implying that all the components are well aligned with the CG as described in Section~\ref{sec:corr}.
The alignments become weaker at large scales.
The shape of the histograms is quite similar among all the components, implying that they are aligned with each other.

\subsection{Radial dependence} \label{sec:statradial}
\begin{figure*}
	\includegraphics[width=1.4\columnwidth]{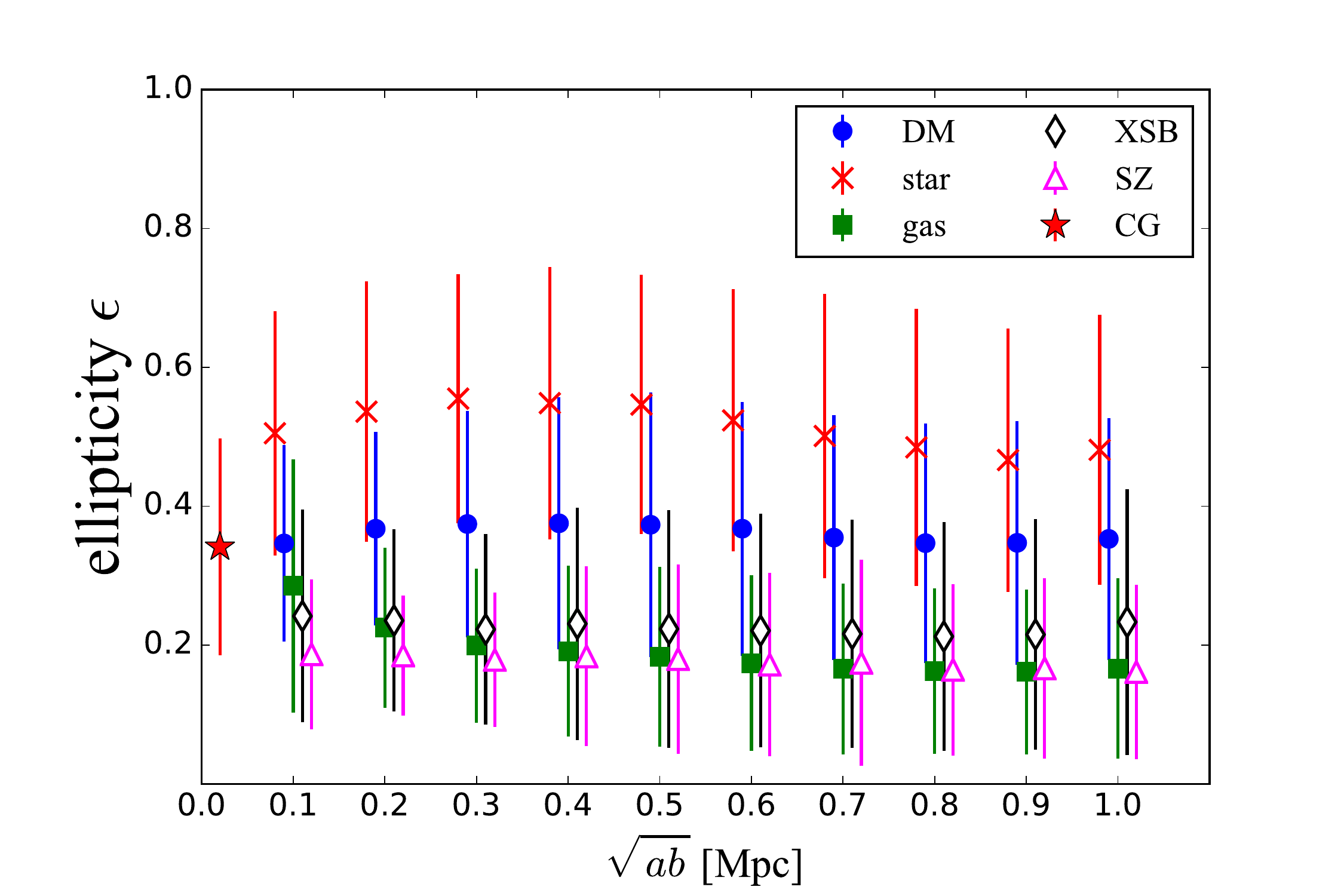}
\caption{The mean ellipticities of different components against $\sqrt{ab}$ of the fitted ellipses. 
The quoted error-bars indicate the corresponding standard deviation.  
  Symbols of DM (filled circles), star (crosses), XSB (diamonds), and SZ (open triangles) are shifted horizontally by 
  $-0.01$, $-0.02$, $0.01$, and $0.02$\,Mpc, respectively just for illustration purpose.
  A red star symbol at $\sqrt{ab}=20$\,kpc represents a mean value of the ellipticity of the CG.
  }
    \label{fig:r-ell}
\end{figure*}
\begin{figure*}
	\includegraphics[width=1.4\columnwidth]{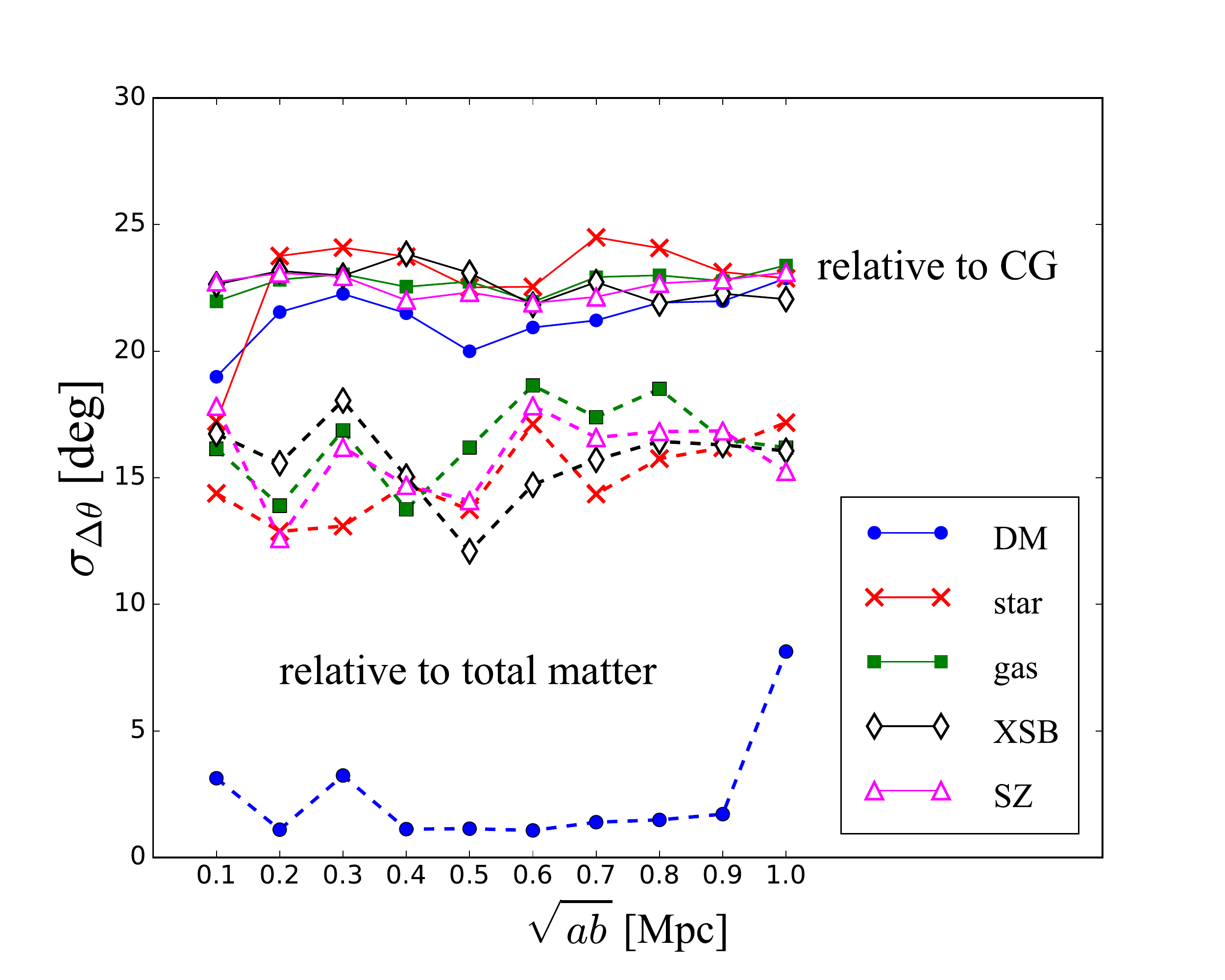}
\caption{
The rms of the position angle difference for different components against $\sqrt{ab}$ of the fitted ellipses. 
Filled circles, crosses, filled squares, open diamonds, and open triangles correspond to the rms values of DM, star, gas, XSB, and SZ, respectively.
Solid and dashed lines indicate the rms of position angle 
relative to the CG and total surface matter density, respectively.
}
    \label{fig:rms}
\end{figure*}
\begin{table}
	\centering
	\caption{Values of mean ellipticities and their errors at $\sqrt{ab}=0.1$, 0.5, and 1.0\,Mpc.}
	\label{tab:ell}
	\begin{tabular}{ l l l l } 
		\hline
		$\sqrt{ab}$ & $0.1$\,Mpc & $0.5$\,Mpc & $1.0$\,Mpc \\
		\hline
		tot     & $0.36\pm0.01$ & $0.36\pm0.02$ & $0.33\pm0.02$\\
		DM    & $0.35\pm0.01$ & $0.37\pm0.02$ & $0.36\pm0.02$\\
		star    & $0.50\pm0.02$ & $0.54\pm0.02$ & $0.48\pm0.02$\\
		gas    & $0.29\pm0.02$ & $0.18\pm0.01$ & $0.17\pm0.01$\\
		XSB   & $0.23\pm0.02$ & $0.23\pm0.02$ & $0.24\pm0.02$\\
		SZ      & $0.19\pm0.01$ & $0.18\pm0.01$ & $0.16\pm0.01$\\
		\hline
	\end{tabular}
\end{table}
Fig.~\ref{fig:r-ell} plots the mean ellipticities of different components against the ellipse scale $\sqrt{ab}$.
Ellipticities for each component are almost constant at all scales
except for that of gas density distribution, which systematically decreases with increasing $\sqrt{ab}$.
This is partly because the position of the CG
is sometimes offset from the potential minimum which corresponds to the density peak of gas components.
Since we fix the centre to the centre of mass of the CG, this mis-centring effect causes the elongation of the gas density ellipse at the most inner part toward the 
direction of gas density peak, resulting in relatively high ellipticities.
Nevertheless, we fix the centre to the CG instead of the potential minimum 
to make it easier to compare our results to those from observations in which the potential minimum is not readily obtained. 

In the outer region, ellipticities of XSB are systematically higher than those of gas and SZ.
XSB is expressed as the integral of the square of the gas number density, $\int n_{\rm gas}^2 {\rm d}l$, 
whereas SZ is computed as $\int n_{\rm gas}T_{\rm gas} {\rm d}l$.
Since mean ellipticities of gas and SZ are similar to each other for outer regions ($\sqrt{ab}>0.3$\,Mpc), 
the temperature distribution is not substantially inhomogeneous.
Thus, the relatively higher values of XSB ellipticities might be caused by the inhomogeneity of the gas density.

Table~\ref{tab:ell} shows mean values of ellipticities and their errors for DM, star, and XSB.
While the quoted error-bars in Fig.~\ref{fig:r-ell} indicate the standard deviations of ellipticities,
the errors in Table~\ref{tab:ell} indicate errors of mean values of ellipticities, which is simply computed by dividing the standard deviations by square root of $N_{\rm cl} = 120$.
These values are consistent with those of \cite{suto17}
within error-bars despite the different method to fit the ellipses (see right panel of Fig.~8 in \citealt{suto17}).
We will compare these values with observations in Section~\ref{sec:obs}.

Fig.~\ref{fig:rms} shows the rms of position angles relative to the CG computed as
\begin{equation}
\sigma^2 _{\Delta\theta, A} \equiv \frac{1}{N_{\rm cl}}\sum^{N_{\rm cl}}_{i=1} (\theta_{A, i}-\theta_{{\rm CG}, i})^2.
\label{eq:rms}
\end{equation}
If the distribution of position angles relative to the CG is perfectly random, the value of rms is expected to be:
\begin{equation}
\sigma_{\Delta\theta, {\rm random}}^2 
= \frac{\displaystyle\int_{0}^{90}  \theta^2 {\rm d}\theta}{\displaystyle\int_{0}^{90} {\rm d}\theta}
= \left( 90/\sqrt{3} \right)^2
\sim ( 52^{\circ} )^2.
\end{equation}
The values of rms for all the components are $20^{\circ}\leq\sigma_{\Delta\theta}\leq25^{\circ}$ and are smaller than $52^{\circ}$ 
at all ellipse scales, indicating that they are well aligned with the major-axis of the CG.
For comparison, \cite{2012JCAP...05..030S} studied position angles between the major-axes of dark matter haloes for inner region and outer region based on $N$-body simulations.
They showed the position angles between those computed from the innermost region, $0.1$ times virial radius of the host halo $r_{\rm vir}$
and those computed from different scales.
A mean value of the position angle is $\sim20^{\circ}$ at $r_{\rm vir}$ (see their Fig.~\textcolor{blue}{7}), which is consistent with our result.

Solid lines in Fig.~\ref{fig:rms} suggest that position angles of the CG are mis-aligned with the other components.
Fig.~\ref{fig:rms} also plots the rms of position angles relative to the total matter density distribution by dashed lines.
The density distribution of dark matter is very significantly aligned with that of total matter,
simply because total matter density distribution is dominated by dark matter distribution.
The rms values for the other components are $10^{\circ}\leq\sigma_{\Delta\theta}\leq20^{\circ}$,
which are systematically smaller than those relative to the CG.
This result indicates that the alignment with the total matter distribution is better than that with the major-axis of the CG.

Stacking analysis is often used to estimate ellipticities of galaxy clusters from weak lensing.
In the stacking analysis, a prior information of position angles of matter density distribution is important to reconstruct the shape of clusters.
There are two proxies of position angles of matter distributions that are adopted in the literature.
One is that of the major-axis of the BCG, and the other is that of the satellite galaxy distribution.
Assuming that
(i)  the CG in the current simulation can be regarded as the BCG in observations, and
(ii) stellar mass density distribution in the current simulation matches luminosity distribution of satellite galaxies, 
our result suggests that the satellite galaxy distribution is a better prior for the stacking analysis than the BCG, at all scales.
Although the satellite distribution is a better prior than the BCG, 
one should keep in mind that there is a non-negligible scatter between position angles of stellar components and total matter distribution,
$\sigma_{\Delta\theta}\sim15^{\circ}$, which must be taken into account when interpreting the stacking analysis results.

\section{Comparison with Observations}\label{sec:obs}
Although the Horizon-AGN simulation is a state-of-the-art cosmological hydrodynamical simulation,
it cannot perfectly reproduce the real universe.
The results described above sections are thus valid only for the specific situation 
such as adopted cosmological parameters, mass resolution, star formation process, feedback process, and so on.
Confronting our theoretical predictions based on the Horizon-AGN simulation should therefore provide a means of testing cosmological models 
as well as baryon physics implemented in the simulation.
Here we tentatively compare our results with available observational data to check the validity of our results.

\subsection{Comparison with ellipticities of observed clusters} \label{sec:obsell}
\begin{table*}
	\centering
	\caption{Values of mean ellipticities and their errors in various observations. SL and WL denotes strong and weak lensing, respectively.
	For weak lensing, a prior information for the major-axis of a cluster is shown if a stacking analysis is used. 
	$N_{\rm obs}$ denotes the number of galaxy clusters used to estimate values of ellipticities.
	We also show the range of radii (scale) used to derive the ellipticities for reference.
	}
	\label{tab:ellobs}
	\begin{threeparttable}
	\begin{tabular}{ l l l l l l  } 
	\hline
	\hline
		Reference & Component (prior) & data set & $N_{\rm obs}$ & scale & ellipticity\\
		\hline
		 \cite{2010ApJ...719.1926K}     & XSB                              & XMM-Newton                & 61        & $0.1-0.4 ~r_{200}$     & $0.21\pm0.004$             \\
		 \cite{2012ApJ...755..116L}       & XSB                              & Chandra and ROSAT   & 31         & $0.04-1~ r_{500}$       & $0.18\pm0.05$               \\
		 \cite{2009ApJ...695.1446E}     & WL (member galaxies) & SDSS                            & 4281    & $0.5-5h^{-1}$\,Mpc     & $0.52^{+0.09}_{-0.14}$   \\
		 \cite{2010MNRAS.404..325R} & SL                                 & HST/Keck                      & 18        & $<250$\,kpc                & $0.30\pm0.13$\tnote{*}   \\	
		\cite{2010MNRAS.405.2215O} & WL                                & Subaru/Supreme-Cam & 18         & $0.1-1.5h^{-1}$\,Mpc  & $0.46\pm0.04$                \\
		\cite{2012MNRAS.420.3213O} & SL                                 & SDSS/Subaru               & 25         & $<100$\,kpc               & $0.38\pm0.05$                \\
		\cite{2012MNRAS.420.3213O} & WL                                & SDSS/Subaru               & 25         & $0.1-3h^{-1}$\,Mpc     & $0.47\pm0.06$                \\
		\cite{2016MNRAS.457.4135C} & WL (BCG)                     & SDSS                            & 2700     & $0.1-4h^{-1}$\,Mpc     & $0.19\pm0.05$\tnote{*}   \\	
		\cite{donahue}                           & XSB                              & Chandra                        & 25         & $500$\,kpc                 & $0.12\pm0.06$                \\
		\cite{donahue}                           & SZ                                 & Bolocam                       & 20         & $500$\,kpc                  & $0.10\pm0.06$               \\
		\cite{donahue}                           & SL/WL                           & HST                              & 25         & $500$\,kpc                  & $0.20\pm0.08$               \\
		\cite{2017MNRAS.467.4131V} & WL (BCG)                      & GAMA/KiDS                 & 2355     & $40-250$\,kpc             & $0.55\pm0.21$\tnote{*}   \\
		\cite{2017MNRAS.467.4131V} & WL (BCG)                      & GAMA/KiDS                 & 2355     & $250-750$\,kpc           & $0.10\pm0.23$\tnote{*}   \\
		\cite{2017MNRAS.467.4131V} & WL (member galaxies)  & GAMA/KiDS                 & 2672     & $40-250$\,kpc             & $-0.08\pm0.20$\tnote{*}   \\
		\cite{2017MNRAS.467.4131V} & WL (member galaxies)  & GAMA/KiDS                 & 2672     & $250-750$\,kpc           & $0.66\pm0.23$\tnote{*}    \\		
		\cite{2018MNRAS.475.2421S} & star                                & SDSS                           & 10428   & $<1$\,Mpc$/h$            & $0.42\pm0.04$                \\
		\cite{2018MNRAS.475.2421S} & WL (member galaxies)  & SDSS                           & 10428   & $0.1-2$\,Mpc$/h$       & $0.45\pm0.09$                \\
		\cite{2018MNRAS.475.2421S} & WL (BCG)                      & SDSS                           & 6681     & $0.1-2$\,Mpc$/h$       & $0.23\pm0.03$               \\
		\hline
	\end{tabular}
	\begin{tablenotes}\footnotesize
	\item[*]{
	Since the definition of the ellipticity in the paper is different from our definition $\epsilon\equiv1-b/a$, 
	we convert the value of ellipticity shown in the paper to our definition. }
	\end{tablenotes}
	\end{threeparttable}
\end{table*}
Table~\ref{tab:ellobs} summarizes various observations of cluster ellipticities, which should be compared with 
our results shown in Fig.~\ref{fig:r-ell} and Table~\ref{tab:ell}.
Below we discuss individual observations listed in Table~\ref{tab:ellobs}.

\cite{2010ApJ...719.1926K} measured the axis ratios of X-ray surface brightness
in the XMM-Newton cluster catalogue compiled by \cite{2008A&A...478..615S}.
Note that the method to fit the ellipse for X-ray image
is based on \cite{1987MNRAS.226..747J} and is different from our method.
The mean values of axis ratios are $0.78$, $0.81$, $0.79$, and $0.78$ at 
$R=0.1$, $0.2$, $0.3$, $0.4r_{200}$, respectively, where $R$ is semi-major axis of ellipses.
The mean value of ellipticities $\epsilon=0.21$ is consistent our result $\epsilon=0.23\pm0.02$ (Table~\ref{tab:ell}) within the error-bar.

\cite{2012ApJ...755..116L} used clusters observed by Chandra and ROSAT
and measured their ellipticities by the tensor method that is similar to our method described in Section~\ref{sec:fit}.
They obtained a mean value of ellipticities, $\epsilon=0.18\pm0.05$, for the local relaxed clusters.
This value is also consistent with our result $\epsilon=0.23\pm0.02$.

Gravitational lensing is a powerful tool to probe the mass distribution of clusters.
The ellipticity has been measured in various studies through both strong and weak lensing methods.
\cite{2009ApJ...695.1446E} analysed 4281 clusters from the catalogue of \cite{2007ApJ...660..239K}
created from Sloan Digital Sky Survey (SDSS) data.
They stacked the weak lensing signals of individual clusters by rotating a cluster to align the major axis of the satellite galaxy distribution.
They corrected systematic effects from anisotropic point spread function (PSF) following \cite{2002AJ....123..583B} and \cite{2003MNRAS.343..459H}.
The errors on the shear map were taken into account by 
\begin{equation}
\sigma^2_{\gamma_{\rm T}} = \sigma^2_{i}  + \sigma^2_{\rm SN}
\end{equation}
where $\sigma_{i}$ includes the shot noise due to the finite number of photons and detector noise,
and $\sigma_{\rm SN}$ denotes the shape noise coming from intrinsic variance of galaxy shapes
(see their equation~\textcolor{blue}{12})

They fitted the stacked signals by an elliptical \cite{1997ApJ...490..493N} profile and obtain the axis ratio $b/a=0.48^{+0.14}_{-0.09}$ 
that corresponds to $\epsilon=0.52^{+0.09}_{-0.14}$.
This ellipticity should be regarded as a lower limit because in stacking they implicitly 
assumed the perfect alignment between major axis of cluster mass distribution and that of satellite galaxy distribution,
which is not the case in our result (see Fig.~\ref{fig:rms}).
Since this misalignment smears out the stacked ellipticity signal, the real value would be slightly larger. 
Assuming our result $\sigma_{\Delta\theta}=15^{\circ}$ as the rms, the ellipticity is expected to be higher by a few percent.
Nevertheless, their value of mean ellipticity is consistent with our result $\epsilon=0.36\pm0.02$ (Table~\ref{tab:ell})
within an error-bar even if the effect of the misalignment is taken into account.

\cite{2010MNRAS.404..325R} measured the ellipticities of clusters taken from Local Cluster Structure Survey.
They fitted strong lensing data with the elliptical mass distribution using LENSTOOL \citep{1993PhDT.......189K, 2007NJPh....9..447J}, 
and obtained averaged ellipticity $\langle\epsilon_{\rm 2D}\rangle=0.34\pm0.14$ in the inner region ($<250$\,kpc).
This value is consistent with our result $\epsilon=0.36\pm0.01$ (Table~\ref{tab:ell}), which may imply that 
the bias described in the paper that strong lensing clusters are expected to be rounder in the sky is not very strong.

\cite{2010MNRAS.405.2215O} reported one of the most significant detections of the cluster ellipticity with gravitational lensing at $7\sigma$ confidence level.
They used weak lensing signals of X-ray luminous clusters from Subaru/Suprime-Cam imaging data \citep{2010PASJ...62..811O}.
They corrected anisotropic PSF following \cite{1995ApJ...449..460K}.
They considered both the intrinsic shape noise of galaxies and cosmic shear due to large scale structure.
They measured ellipticities for individual clusters without any prior by directly comparing the lensing shear map with elliptical model predictions,
and obtained the mean ellipticity $\langle\epsilon\rangle=0.46\pm0.04$.
This value is higher than our result of $\epsilon=0.36\pm0.02$, presumably because of the higher cluster masses ($M\sim10^{15}M_{\odot}$) of these clusters.
Many studies suggested that dark matter haloes with higher masses have higher ellipticities
\citep[e.g.][]{2005ApJ...629..781K, 2006MNRAS.366.1503P, 2007ApJ...664..117G, 2007MNRAS.377..883F, 2014MNRAS.443.3208D}

\cite{2012MNRAS.420.3213O} obtained the similar value of ellipticity $\langle\epsilon\rangle=0.47\pm0.06$ for strong lensing galaxy clusters from SDSS.
They took into account anisotropic PSF and noise following \cite{2010MNRAS.405.2215O}.
They analysed their weak lensing signals through the stacking analysis by using position angles derived from strong lensing analysis as a prior information.
They claimed that this prior enables much more robust stacking analysis than using other priors.
This prior is however only available for the strong lensing clusters.
They also modeled these clusters by using strong lensing method described in \cite{2009ApJ...699.1038O} and \cite{2010PASJ...62.1017O},
and found noisy but slightly lower mean ellipticity $\langle\epsilon\rangle=0.38\pm0.05$.

\cite{2016MNRAS.457.4135C} used the technique to measure the quadrupole weak lensing signal, and applied it to a sample of SDSS clusters.
They corrected anisotropic PSF following \cite{2012MNRAS.425.2610R} and \cite{2014MNRAS.440.1296H}.
They considered the noise from the intrinsic shape of galaxies and measurement on each background galaxy following \cite{2013MNRAS.432.1544M} and \cite{2012ApJS..201...32S}.
They obtained the best fit value of the mean ellipticity of $\epsilon=0.19$ with $1\sigma$ uncertainty of $\sim0.05$.
They ascribed this smaller value to the misalignment between major axis of the BCG and that of cluster halo which is implicitly assumed to be aligned.
Given the large uncertainty, their result is broadly consistent with our result. 

\cite{2018MNRAS.475.2421S} applied the quadrupole technique to SDSS clusters.
They estimated anisotropic PSF, measurement noise and noise from the intrinsic shape following \cite{2016MNRAS.457.4135C}. 
The resulting mean ellipticity value with a prior of the satellite galaxy distribution 
is $\langle\epsilon\rangle=0.45\pm0.09$ after correcting for Poisson sampling.
They also measured the ellipticity of satellite galaxy distribution as $\langle\epsilon\rangle=0.42\pm0.04$,
and that derived from stacked weak lensing with a prior of the CG major axis as $\langle\epsilon\rangle=0.25\pm0.06$.
By comparing these ellipticity values, they also estimated the rms misalignment angle of 
$30^{\circ}$ between the CG and dark matter halo and 
$18^{\circ}$ between satellite galaxies and dark matter halo.
These misalignment values are in good agreement with our result (see also Section~\ref{sec:obspa}).

\cite{2017MNRAS.467.4131V} used an estimator similar to \cite{2016MNRAS.457.4135C} and constrained the average ellipticity of galaxy groups obtained from 
Galaxy And Mass Assembly (GAMA) survey combined with the weak lensing signal 
measured by \cite{2017MNRAS.465.1454H} from the Kilo Degree Survey (KiDS).
They did not consider anisotropic PSF but consider the intrinsic shape noise.
They compared different priors for stacking analysis of weak lensing signals at different scales.
Their resulting values of the mean ellipticity are 
$\epsilon=0.38\pm0.12$ ($40$\,kpc$<R<250$\,kpc) and 
$\epsilon=0.05\pm0.13$ ($250$\,kpc$<R<750$\,kpc) for the BCG prior,
whereas $\epsilon=-0.04\pm0.11$ and $\epsilon=0.349\pm0.13$, respectively for the prior of the satellite galaxy distribution.
They concluded that the BCG major-axis (satellite galaxy distribution) is aligned (misaligned)
with the dark matter halo orientation on small scales ($<250$\,kpc)
whereas the BCG major-axis (satellite galaxy distribution)
is misaligned (aligned) with dark matter on large scales ($>250$\,kpc).
This result appears to be inconsistent with our result which indicates that the distribution of satellite galaxies 
is aligned better than the major-axis of the CG at {\it all} scales, $100-1000$\,kpc.
This discrepancies are partly because they use galaxy groups with $M_{200}\sim10^{13}M_{\odot}$
rather than galaxy clusters we considered in this paper, $M_{200}\sim10^{14}M_{\odot}$.
Nevertheless, further work is needed to explain this inconsistency,
for example by analyzing the galaxy groups with masses of $M_{200}\sim10^{13}M_{\odot}$ in the Horizon-AGN simulation.

\cite{donahue} systematically measured the ellipticities of X-ray surface brightness, Sunyaev-Zel'dovich effect (SZE), 
gravitational lensing map, and the BCG
for clusters from Cluster Lensing and Supernova survey with Hubble Space Telescope (CLASH).
They used X-ray data from {\it Chandra X-ray Observatory} and measured the axis ratio based on the procedure described in \cite{2015ApJ...805..177D}.
The method is almost the same as the one we used (see Section~\ref{sec:fit}).
The same procedure was applied to the SZ Compton $y$-parameter map obtained from the Bolocam SZ images \citep[see ][]{2013ApJ...768..177S,  2015ApJ...806...18C}.
They found the mean axis ratios $0.09\pm0.05$ and $0.1\pm0.06$ for XSB and SZ, respectively at scales of $500$\,kpc.
These values are much lower than our results, which is not surprising because their clusters were selected to be nearly circular in X-ray.
They also measured the ellipticity of gravitational lensing surface mass density map created from both strong and weak lensing.
The detail of the lensing analysis is described in \cite{2015ApJ...801...44Z}.
The resulting mean ellipticity value is $0.2\pm0.08$ at $500$\,kpc, and is also lower than our result ($\epsilon=0.36\pm0.02$) probably due to the selection effect.

Strictly speaking, the observational ellipticities derived from lensing analysis are not exactly the same as
those of total matter distributions in the simulation
since observable in the lensing is shear signals whereas the total matter distributions correspond to the convergence signals.
In addition, observations have various systematics such as Poisson noise of background galaxies, intrinsic alignment, and contamination of point spread function. 
The most straightforward way to compare our results with these observations is to create mock shear catalogue and 
evaluate the ellipticities by adopting the same lensing method.
Such analysis is beyond the scope of this paper, and will be presented elsewhere.

\subsection{Comparison with observed position angle distributions} \label{sec:obspa}
\begin{figure*}
	\includegraphics[width=2\columnwidth]{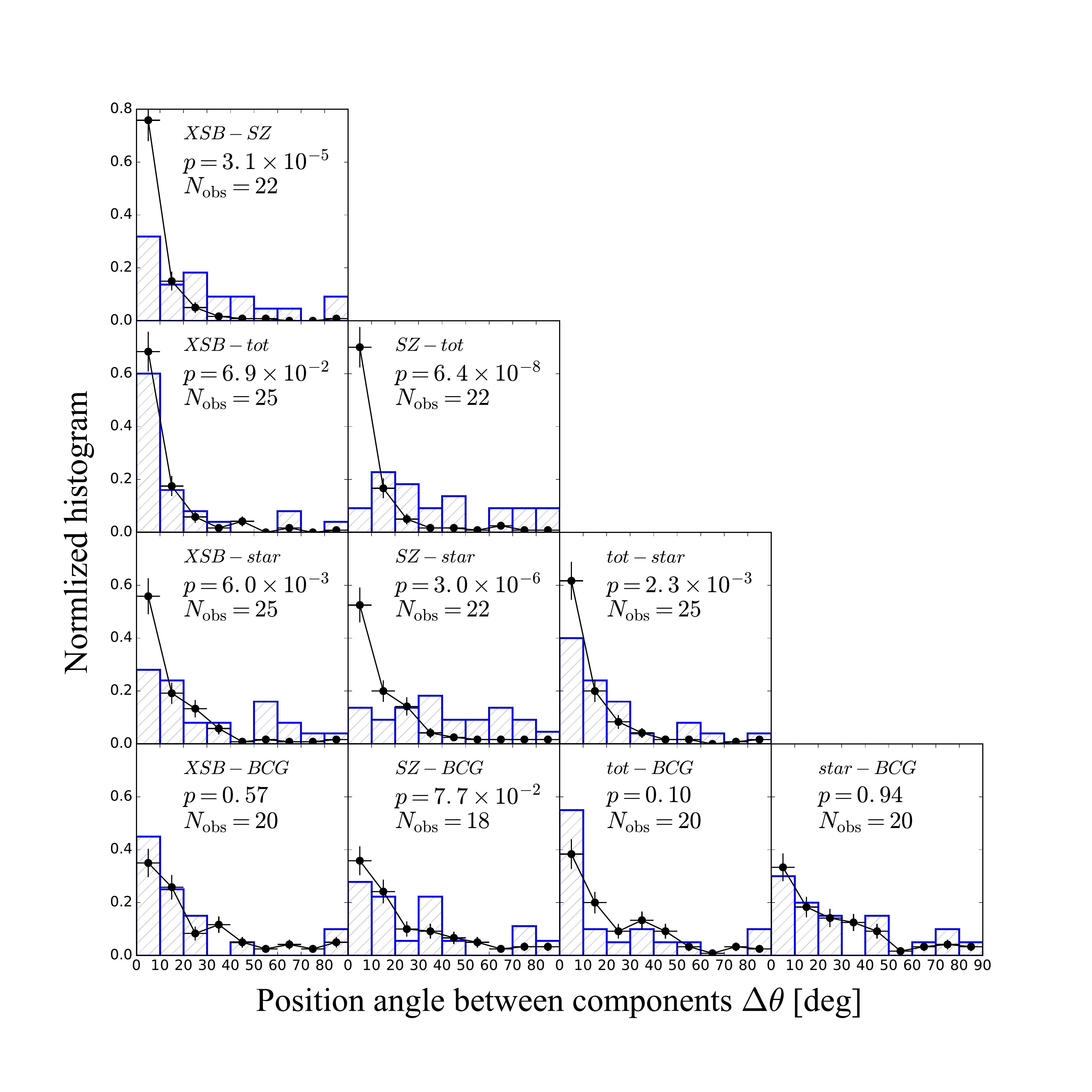}
    \caption{Comparison of observed distribution of relative position angles of different components 
    against our simulated data. 
    Blue hatched histograms and black symbols are normalized histograms of relative position angles in observations and our simulation, respectively.
    $N_{\rm obs}$ indicates number of galaxy clusters which both components are available to estimate position angles,
    and $p$ denotes $p$-values of the Kolmogorov-Smirnov test.
    }
    \label{fig:obs}
\end{figure*}
In this subsection, we regard the CG of the simulation as the BCG in observations,
since the CG is supposed to be almost identical to the observed BCG as described in Section~\ref{sec:cluster}.

Fig.~\ref{fig:obs} compares normalized histograms of position angles between two components from the Horizon-AGN simulation with observations.
The observational data are based on position angles of the BCG from \cite{2015ApJ...805..177D}, those of XSB, SZ, and tot from \cite{donahue},
and those of stellar distribution from \cite{2017NatAs...1E.157W}.
We choose 25 clusters in \cite{donahue} (see their Table~\textcolor{blue}{1} and Fig.~\textcolor{blue}{1}).
Twenty of these clusters were selected based on their relatively round X-ray shape and with prominent BCG at their centre being well aligned with X-ray.
For these 20 clusters, \cite{2015ApJ...805..177D} measured the position angles of BCGs
by using the surface brightness weighted tensor method.
They obtained the position angles for both ultraviolet and near-infrared data. 
We use those derived from near-infrared data because near-infrared light is dominated by old stars
which is expected to dominate the mass in the centre of BCG.
The values are summarized in Table~\textcolor{blue}{3} in \cite{2015ApJ...805..177D}.
\cite{donahue} measured the position angles of XSB and SZ by using the same method as \cite{2015ApJ...805..177D},
and those of total matter distributions by utilizing the otherwise identical procedures for lensing-based surface mass density maps.
We use their values estimated within $500$\,kpc.
The values for XSB, SZ, and tot are summarized in Table~\textcolor{blue}{3}, \textcolor{blue}{6}, and \textcolor{blue}{5} in \cite{donahue}, respectively.
\cite{2017NatAs...1E.157W} measured position angles of the member galaxy distribution by computing the moments of inertia 
of the red sequence galaxy distribution.
Table~\textcolor{blue}{1} in \cite{2017NatAs...1E.157W} summarizes the resulting values.

We compare position angle between these components 
(10 combinations for 5 different components mentioned above) with our result shown in Fig.~\ref{fig:pdf}.
We use our measurement at $500$\,kpc following \cite{donahue}.
Fig.~\ref{fig:obs} shows the resulting $p$-values of the Kolmogorov-Smirnov test.
We find that histograms from the simulation generally agree well with observations, except for those related to SZ.
One of the reason of the low $p$-values related to the SZ is the poor angular resolution 
of Bolocam with a full width half maximum of $58''$, which makes measurement of the position angles for the SZ maps very noisy.
The observed distribution would be more consistent with our simulation result once the measurement errors of the position angles are taken into account.

The relatively lower $p$-values related to the stellar distributions are partly because there are not sufficient numbers of member galaxies,
and therefore Poisson noise affects the position angle measurement.
To draw more robust conclusion, we have to take into account of selection effects and differences in measurement methods.
Nevertheless, broad agreements between the simulation and the observations are encouraging, which invites more careful analysis of observational data 
based on our simulation results.

While \cite{2017MNRAS.472.1163C} focused on the three-dimensional alignment angle between galaxies and their host dark matter haloes,
they also calculated projected shapes for galaxies in Horizon-AGN simulation and matched dark matter haloes in Horizon-DM simulation.
They compared the major axises of galaxies in Horizon-AGN simulation and those of matched dark matter haloes in Horizon-DM simulation
and derived the alignment angle distribution (see their Fig.~\textcolor{blue}{B1}).
They obtained a mean alignment angle and dispersion of $-2^{\circ}\pm48^{\circ}$, which is marginally consistent with our result of $5^{\circ}\pm30^{\circ}$ (Fig.~\ref{fig:pdf})
though both galaxies and host dark matter haloes in our analysis are in Horizon-AGN simulation.
\cite{2015MNRAS.453..469T} analysed the shapes and position angles of stellar and dark matter haloes in the MassiveBlack-II simulation \citep{2015MNRAS.450.1349K},
which is a cosmological hydrodynamical simulation including stellar and AGN feedback in
a volume of $(100h^{-1} {\rm Mpc})^3$ comparable to that of the current Horizon simulation.
They obtained a mean projected position angle between galaxies and dark matter haloes of $11^{\circ}$ (see their Table~\textcolor{blue}{2}),
which is smaller than our result of $21^{\circ}$ (Fig.~\ref{fig:pdf}).
The detailed comparison, however, is difficult since different method is used to derive the position angles.
\cite{2015MNRAS.453..721V} reported the shapes and position angles of dark matter, stellar, and gas components in the EAGLE \citep{2015MNRAS.446..521S} 
and cosmo-OWLS \citep{2014MNRAS.441.1270L} simulations, which are smoothed particle hydrodynamics simulations \citep{1992ARA&A..30..543M}. 
They obtained median position angles between stellar and total matter components of $10^{\circ}-25^{\circ}$ (see their Fig.~\textcolor{blue}{13}),
which is consistent with our result of $10^{\circ}$ (Fig.~\ref{fig:pdf}).

\section{Conclusions}\label{sec:summary}
We have a presented comprehensive study of projected alignments of stellar, gas, and dark matter distributions in galaxy clusters based on 
the ellipse fit to projected images of 40 galaxy clusters with masses larger than $5\times10^{13}M_{\odot}$
from the Horizon-AGN cosmological hydrodynamical simulation.
For each cluster, we consider six different components; X-ray surface brightness, Compton $y$-parameter of Sunyaev-Zel'dovich effect,
total, dark matter, stellar, and gas surface mass density distributions, which can be compared with observations.
For each cluster we consider three different projection directions and regard them as independent. 
We thus have 120 independent projected images for each component, which allow statistical studies of projected
non-sphericities and alignments between different components.

We have applied the tensor method to these images at 10 different scales over the range $\sqrt{ab}=0.1-1.0$\,Mpc to derive the ellipticities and position angles at each scale.
We also measured the ellipticities and position angles for central galaxies (CG), which are selected by the most massive galaxy in each cluster.
Our main results are summarized as follows.

(i) Projected distributions of dark matter, stellar, and gas components are well aligned with the major axis of the CG, 
with the root mean square of their position angle differences of 
$\sigma_{\Delta\theta}\equiv|\theta-\theta_{\rm CG}| \sim 20^{\circ}-25^{\circ}$, which is nearly independent of scales and components.
 
(ii) Projected distributions of dark matter, stellar, and gas components are aligned with total matter density distribution better than with the CG,
with $\sigma_{\Delta\theta}$ of $1^{\circ}-2^{\circ}$ for DM, and $10^{\circ}-20^{\circ}$ between total matter and the other components.

(iii) 
Ellipticities of all the components do not show tight correlation with that of the CG
even if position angles are fairly aligned with each other, $\Delta\theta\leq10^{\circ}$.

(iv)
Ellipticities of dark matter, stellar, and gas distributions, correlate with each other.
The correlation is stronger for DM-star, DM-gas, and star-gas in this order.
The strongest correlation between dark matter and stellar components is simply because old stars that dominate stellar masses of clusters follow
dark matter distribution fairly well.
The better correlation between dark matter and gas components reflects the fact that gas components follow 
the gravitational potential of the cluster for which the contribution from dark matter dominates.

(v)
Values of mean ellipticities and distributions of position angles derived in this paper are broadly consistent with various observations.

The stronger alignment of stellar mass components relative to the total matter density than the CG implies that 
the distribution of satellite galaxy is a better prior information for the stacking analysis of weak lensing than the major axis of the CG.
Our result indicates that the alignment is not perfect and hence 
the rms of position angle differences should be taken into account to correctly interpret the elliptical signals of weak lensing from stacking analysis.
In this paper we have derived quantitative estimates of the rms values, which should provide useful guidance for interpreting the lensing signal from stacking analysis.
Our result also suggests that X-ray surface brightness and SZ serve as useful prior for the stacking analysis, 
since alignments between XSB or SZ and total matter distributions are as tight as those between star and total matter distributions in our result.
To summarize, our comprehensive analysis of alignments
in the realistic hydrodynamical simulation provides useful clue in interpreting various stacking observations 
as well as designing future observations along this line. 
Our result may also be useful for studying correlations between cluster shapes and surrounding
matter distributions up to very large scales \citep[e.g.][]{2017arXiv171200094O, 2018MNRAS.474.1165P}.

We still find that the position angles of matter distributions in galaxy clusters
are moderately aligned with the major-axis of the central galaxy up to $\sim 1$\,Mpc.
Numerical studies based on $N$-body simulation reported the similar alignment between dark matter halo at inner region and those at outer regions \cite[e.g.][]{2012JCAP...05..030S}.
In order to investigate the origin of the alignment, we will work on the evolution of galaxy clusters and central galaxy in the Horizon-AGN simulation, which will be reported elsewhere.

\section*{Acknowledgements}
We warmly thank the Horizon-AGN team, especially Y. Dubois, C. Pichon and J. Devriendt, for making their simulation available.
T.O. and Y.S. gratefully acknowledge the hospitality of Observatoire de la C\^ote d'Azur. 
We thank an anonymous referee for useful comments.
T.O. is supported by Advanced Leading Graduate Course for Photon Science (ALPS) at the University of Tokyo.
This work is supported partly by JSPS Core-to-Core Program “International Network of Planetary Sciences,”
This work is supported in part by Japan Society for the Promotion of Science (JSPS) KAKENHI Grant Number 
17J05056 (T.O.), 17K14273 (T.N.), JP26800093 (M.O.), JP15H05892 (M.O.), 25400236 (T.K.), and 24340035 (Y. S.).
T.N. acknowledges Japan Science and Technology Agency (JST) CREST Grant Number JPMJCR1414.


Note added in proof: After this paper was submitted, three papers relating to the ellipticities and position angles of galaxy clusters were posted on the arXiv
\citep{2018arXiv180400664U, 2018arXiv180400667S, 2018arXiv180400676C}.
These three papers analyse 20 galaxy clusters from the CLASH survey.
They measure both two and three dimensional shapes of mass distributions by combining 
the X-ray and gravitational lensing data.
\cite{2018arXiv180400664U} estimates the median projected axis ratio of $0.67\pm0.07$,
corresponding $\epsilon=0.33\pm0.07$.
The median value is lower than our resulting value $\epsilon0.36\pm0.02$ (see Table~\ref{tab:ell}) 
due to selection effect.
This result is not unnatural because they select the CLASH clusters which have circular shapes in X-ray as also discussed in section~\ref{sec:obsell}.
They also evaluate the misalignment angles of baryonic components (X-ray, thermal Sunyaev-Zel'dovich effect, brightest cluster galaxy) 
with respect to the weak lensing.
They conclude that the major-axis of X-ray shows best aligned with mass distribution derived from weak lensing with a median misalignment angle of $21^{\circ}\pm7^{\circ}$
(see their Fig.~\textcolor{blue}{6}).
This result is quantitatively consistent with our result which indicates X-ray is aligned better than brightest cluster galaxy with respect to total mass distribution.
The worse alignment of thermal SZ effect with respect to the total mass distribution might be due to large PSF of Bolocam images.

\bibliographystyle{mnras}
\bibliography{reference} 

\begin{thebibliography}{}
\makeatletter
\relax
\def\mn@urlcharsother{\let\do\@makeother \do\$\do\&\do\#\do\^\do\_\do\%\do\~}
\def\mn@doi{\begingroup\mn@urlcharsother \@ifnextchar [ {\mn@doi@}
  {\mn@doi@[]}}
\def\mn@doi@[#1]#2{\def\@tempa{#1}\ifx\@tempa\@empty \href
  {http://dx.doi.org/#2} {doi:#2}\else \href {http://dx.doi.org/#2} {#1}\fi
  \endgroup}
\def\mn@eprint#1#2{\mn@eprint@#1:#2::\@nil}
\def\mn@eprint@arXiv#1{\href {http://arxiv.org/abs/#1} {{\tt arXiv:#1}}}
\def\mn@eprint@dblp#1{\href {http://dblp.uni-trier.de/rec/bibtex/#1.xml}
  {dblp:#1}}
\def\mn@eprint@#1:#2:#3:#4\@nil{\def\@tempa {#1}\def\@tempb {#2}\def\@tempc
  {#3}\ifx \@tempc \@empty \let \@tempc \@tempb \let \@tempb \@tempa \fi \ifx
  \@tempb \@empty \def\@tempb {arXiv}\fi \@ifundefined
  {mn@eprint@\@tempb}{\@tempb:\@tempc}{\expandafter \expandafter \csname
  mn@eprint@\@tempb\endcsname \expandafter{\@tempc}}}

\bibitem[\protect\citeauthoryear{{Allen}, {Schmidt}, {Ebeling}, {Fabian}  \&
  {van Speybroeck}}{{Allen} et~al.}{2004}]{2004MNRAS.353..457A}
{Allen} S.~W.,  {Schmidt} R.~W.,  {Ebeling} H.,  {Fabian} A.~C.,   {van
  Speybroeck} L.,  2004, \mn@doi [\mnras] {10.1111/j.1365-2966.2004.08080.x},
  \href {http://adsabs.harvard.edu/abs/2004MNRAS.353..457A} {353, 457}

\bibitem[\protect\citeauthoryear{{Allen}, {Rapetti}, {Schmidt}, {Ebeling},
  {Morris}  \& {Fabian}}{{Allen} et~al.}{2008}]{2008MNRAS.383..879A}
{Allen} S.~W.,  {Rapetti} D.~A.,  {Schmidt} R.~W.,  {Ebeling} H.,  {Morris}
  R.~G.,   {Fabian} A.~C.,  2008, \mn@doi [\mnras]
  {10.1111/j.1365-2966.2007.12610.x}, \href
  {http://adsabs.harvard.edu/abs/2008MNRAS.383..879A} {383, 879}

\bibitem[\protect\citeauthoryear{{Argyres}, {Groth}, {Peebles}  \&
  {Struble}}{{Argyres} et~al.}{1986}]{1986AJ.....91..471A}
{Argyres} P.~C.,  {Groth} E.~J.,  {Peebles} P.~J.~E.,   {Struble} M.~F.,  1986,
  \mn@doi [\aj] {10.1086/114025}, \href
  {http://adsabs.harvard.edu/abs/1986AJ.....91..471A} {91, 471}

\bibitem[\protect\citeauthoryear{{Aubert}, {Pichon}  \& {Colombi}}{{Aubert}
  et~al.}{2004}]{2004MNRAS.352..376A}
{Aubert} D.,  {Pichon} C.,   {Colombi} S.,  2004, \mn@doi [\mnras]
  {10.1111/j.1365-2966.2004.07883.x}, \href
  {http://adsabs.harvard.edu/abs/2004MNRAS.352..376A} {352, 376}

\bibitem[\protect\citeauthoryear{{Azzaro}, {Patiri}, {Prada}  \&
  {Zentner}}{{Azzaro} et~al.}{2007}]{2007MNRAS.376L..43A}
{Azzaro} M.,  {Patiri} S.~G.,  {Prada} F.,   {Zentner} A.~R.,  2007, \mn@doi
  [\mnras] {10.1111/j.1745-3933.2007.00282.x}, \href
  {http://adsabs.harvard.edu/abs/2007MNRAS.376L..43A} {376, L43}

\bibitem[\protect\citeauthoryear{{Beckmann} et~al.,}{{Beckmann}
  et~al.}{2017}]{2017MNRAS.472..949B}
{Beckmann} R.~S.,  et~al., 2017, \mn@doi [\mnras] {10.1093/mnras/stx1831},
  \href {http://ads.nao.ac.jp/abs/2017MNRAS.472..949B} {472, 949}

\bibitem[\protect\citeauthoryear{{Benson} et~al.,}{{Benson}
  et~al.}{2013}]{2013ApJ...763..147B}
{Benson} B.~A.,  et~al., 2013, \mn@doi [\apj] {10.1088/0004-637X/763/2/147},
  \href {http://adsabs.harvard.edu/abs/2013ApJ...763..147B} {763, 147}

\bibitem[\protect\citeauthoryear{{Bernstein} \& {Jarvis}}{{Bernstein} \&
  {Jarvis}}{2002}]{2002AJ....123..583B}
{Bernstein} G.~M.,  {Jarvis} M.,  2002, \mn@doi [\aj] {10.1086/338085}, \href
  {http://adsabs.harvard.edu/abs/2002AJ....123..583B} {123, 583}

\bibitem[\protect\citeauthoryear{{Bett}, {Eke}, {Frenk}, {Jenkins}  \&
  {Okamoto}}{{Bett} et~al.}{2010}]{2010MNRAS.404.1137B}
{Bett} P.,  {Eke} V.,  {Frenk} C.~S.,  {Jenkins} A.,   {Okamoto} T.,  2010,
  \mn@doi [\mnras] {10.1111/j.1365-2966.2010.16368.x}, \href
  {http://adsabs.harvard.edu/abs/2010MNRAS.404.1137B} {404, 1137}

\bibitem[\protect\citeauthoryear{{Biernacka}, {Panko}, {Bajan}, {God{\l}owski}
  \& {Flin}}{{Biernacka} et~al.}{2015}]{2015ApJ...813...20B}
{Biernacka} M.,  {Panko} E.,  {Bajan} K.,  {God{\l}owski} W.,   {Flin} P.,
  2015, \mn@doi [\apj] {10.1088/0004-637X/813/1/20}, \href
  {http://adsabs.harvard.edu/abs/2015ApJ...813...20B} {813, 20}

\bibitem[\protect\citeauthoryear{{Binggeli}}{{Binggeli}}{1982}]{1982A&A...107..338B}
{Binggeli} B.,  1982, \aap, \href
  {http://ads.nao.ac.jp/abs/1982A%26A...107..338B} {107, 338}

\bibitem[\protect\citeauthoryear{{Bogdan}, {Lovisari}, {Volonteri}  \&
  {Dubois}}{{Bogdan} et~al.}{2017}]{2017arXiv171109900B}
{Bogdan} A.,  {Lovisari} L.,  {Volonteri} M.,   {Dubois} Y.,  2017, preprint,
  \href {http://adsabs.harvard.edu/abs/2017arXiv171109900B} {} (\mn@eprint
  {arXiv} {1711.09900})

\bibitem[\protect\citeauthoryear{{B{\"o}hringer}, {Chon}  \&
  {Collins}}{{B{\"o}hringer} et~al.}{2014}]{2014A&A...570A..31B}
{B{\"o}hringer} H.,  {Chon} G.,   {Collins} C.~A.,  2014, \mn@doi [\aap]
  {10.1051/0004-6361/201323155}, \href
  {http://adsabs.harvard.edu/abs/2014A%26A...570A..31B} {570, A31}

\bibitem[\protect\citeauthoryear{{Bonamente}, {Joy}, {LaRoque}, {Carlstrom},
  {Reese}  \& {Dawson}}{{Bonamente} et~al.}{2006}]{2006ApJ...647...25B}
{Bonamente} M.,  {Joy} M.~K.,  {LaRoque} S.~J.,  {Carlstrom} J.~E.,  {Reese}
  E.~D.,   {Dawson} K.~S.,  2006, \mn@doi [\apj] {10.1086/505291}, \href
  {http://adsabs.harvard.edu/abs/2006ApJ...647...25B} {647, 25}

\bibitem[\protect\citeauthoryear{{Brainerd}}{{Brainerd}}{2005}]{2005ApJ...628L.101B}
{Brainerd} T.~G.,  2005, \mn@doi [\apjl] {10.1086/432713}, \href
  {http://adsabs.harvard.edu/abs/2005ApJ...628L.101B} {628, L101}

\bibitem[\protect\citeauthoryear{{Bryan}, {Kay}, {Duffy}, {Schaye}, {Dalla
  Vecchia}  \& {Booth}}{{Bryan} et~al.}{2013}]{2013MNRAS.429.3316B}
{Bryan} S.~E.,  {Kay} S.~T.,  {Duffy} A.~R.,  {Schaye} J.,  {Dalla Vecchia} C.,
    {Booth} C.~M.,  2013, \mn@doi [\mnras] {10.1093/mnras/sts587}, \href
  {http://adsabs.harvard.edu/abs/2013MNRAS.429.3316B} {429, 3316}

\bibitem[\protect\citeauthoryear{{Chen} et~al.,}{{Chen}
  et~al.}{2015}]{2015MNRAS.454.3341C}
{Chen} Y.-C.,  et~al., 2015, \mn@doi [\mnras] {10.1093/mnras/stv2260}, \href
  {http://adsabs.harvard.edu/abs/2015MNRAS.454.3341C} {454, 3341}

\bibitem[\protect\citeauthoryear{{Cheung} et~al.,}{{Cheung}
  et~al.}{2016}]{2016Natur.533..504C}
{Cheung} E.,  et~al., 2016, \mn@doi [\nat] {10.1038/nature18006}, \href
  {http://adsabs.harvard.edu/abs/2016Natur.533..504C} {533, 504}

\bibitem[\protect\citeauthoryear{{Chisari} et~al.,}{{Chisari}
  et~al.}{2015}]{2015MNRAS.454.2736C}
{Chisari} N.,  et~al., 2015, \mn@doi [\mnras] {10.1093/mnras/stv2154}, \href
  {http://ads.nao.ac.jp/abs/2015MNRAS.454.2736C} {454, 2736}

\bibitem[\protect\citeauthoryear{{Chisari} et~al.,}{{Chisari}
  et~al.}{2016}]{2016MNRAS.461.2702C}
{Chisari} N.,  et~al., 2016, \mn@doi [\mnras] {10.1093/mnras/stw1409}, \href
  {http://ads.nao.ac.jp/abs/2016MNRAS.461.2702C} {461, 2702}

\bibitem[\protect\citeauthoryear{{Chisari} et~al.,}{{Chisari}
  et~al.}{2017}]{2017MNRAS.472.1163C}
{Chisari} N.~E.,  et~al., 2017, \mn@doi [\mnras] {10.1093/mnras/stx1998}, \href
  {http://adsabs.harvard.edu/abs/2017MNRAS.472.1163C} {472, 1163}

\bibitem[\protect\citeauthoryear{{Chiu}, {Umetsu}, {Sereno}, {Ettori},
  {Meneghetti}, {Merten}, {Sayers}  \& {Zitrin}}{{Chiu}
  et~al.}{2018}]{2018arXiv180400676C}
{Chiu} I.,  {Umetsu} K.,  {Sereno} M.,  {Ettori} S.,  {Meneghetti} M.,
  {Merten} J.,  {Sayers} J.,   {Zitrin} A.,  2018, preprint, \href
  {http://adsabs.harvard.edu/abs/2018arXiv180400676C} {} (\mn@eprint {arXiv}
  {1804.00676})

\bibitem[\protect\citeauthoryear{{Clampitt} \& {Jain}}{{Clampitt} \&
  {Jain}}{2016}]{2016MNRAS.457.4135C}
{Clampitt} J.,  {Jain} B.,  2016, \mn@doi [\mnras] {10.1093/mnras/stw254},
  \href {http://adsabs.harvard.edu/abs/2016MNRAS.457.4135C} {457, 4135}

\bibitem[\protect\citeauthoryear{{Czakon} et~al.,}{{Czakon}
  et~al.}{2015}]{2015ApJ...806...18C}
{Czakon} N.~G.,  et~al., 2015, \mn@doi [\apj] {10.1088/0004-637X/806/1/18},
  \href {http://adsabs.harvard.edu/abs/2015ApJ...806...18C} {806, 18}

\bibitem[\protect\citeauthoryear{{Dav{\'e}}, {Spergel}, {Steinhardt}  \&
  {Wandelt}}{{Dav{\'e}} et~al.}{2001}]{2001ApJ...547..574D}
{Dav{\'e}} R.,  {Spergel} D.~N.,  {Steinhardt} P.~J.,   {Wandelt} B.~D.,  2001,
  \mn@doi [\apj] {10.1086/318417}, \href
  {http://adsabs.harvard.edu/abs/2001ApJ...547..574D} {547, 574}

\bibitem[\protect\citeauthoryear{{Despali} \& {Vegetti}}{{Despali} \&
  {Vegetti}}{2017}]{2017MNRAS.469.1997D}
{Despali} G.,  {Vegetti} S.,  2017, \mn@doi [\mnras] {10.1093/mnras/stx966},
  \href {http://adsabs.harvard.edu/abs/2017MNRAS.469.1997D} {469, 1997}

\bibitem[\protect\citeauthoryear{{Despali}, {Giocoli}  \& {Tormen}}{{Despali}
  et~al.}{2014}]{2014MNRAS.443.3208D}
{Despali} G.,  {Giocoli} C.,   {Tormen} G.,  2014, \mn@doi [\mnras]
  {10.1093/mnras/stu1393}, \href
  {http://adsabs.harvard.edu/abs/2014MNRAS.443.3208D} {443, 3208}

\bibitem[\protect\citeauthoryear{{Despali}, {Giocoli}, {Angulo}, {Tormen},
  {Sheth}, {Baso}  \& {Moscardini}}{{Despali}
  et~al.}{2016}]{2016MNRAS.456.2486D}
{Despali} G.,  {Giocoli} C.,  {Angulo} R.~E.,  {Tormen} G.,  {Sheth} R.~K.,
  {Baso} G.,   {Moscardini} L.,  2016, \mn@doi [\mnras]
  {10.1093/mnras/stv2842}, \href
  {http://adsabs.harvard.edu/abs/2016MNRAS.456.2486D} {456, 2486}

\bibitem[\protect\citeauthoryear{{Despali}, {Giocoli}, {Bonamigo}, {Limousin}
  \& {Tormen}}{{Despali} et~al.}{2017}]{2017MNRAS.466..181D}
{Despali} G.,  {Giocoli} C.,  {Bonamigo} M.,  {Limousin} M.,   {Tormen} G.,
  2017, \mn@doi [\mnras] {10.1093/mnras/stw3129}, \href
  {http://adsabs.harvard.edu/abs/2017MNRAS.466..181D} {466, 181}

\bibitem[\protect\citeauthoryear{{Djorgovski}}{{Djorgovski}}{1987}]{1987nngp.proc..227D}
{Djorgovski} S.~G.,  1987, in {Faber} S.~M.,  ed., Nearly Normal Galaxies. From
  the Planck Time to the Present. pp 227--233

\bibitem[\protect\citeauthoryear{{Donahue} et~al.,}{{Donahue}
  et~al.}{2015}]{2015ApJ...805..177D}
{Donahue} M.,  et~al., 2015, \mn@doi [\apj] {10.1088/0004-637X/805/2/177},
  \href {http://adsabs.harvard.edu/abs/2015ApJ...805..177D} {805, 177}

\bibitem[\protect\citeauthoryear{{Donahue} et~al.,}{{Donahue}
  et~al.}{2016}]{donahue}
{Donahue} M.,  et~al., 2016, \mn@doi [\apj] {10.3847/0004-637X/819/1/36}, \href
  {http://adsabs.harvard.edu/abs/2016ApJ...819...36D} {819, 36}

\bibitem[\protect\citeauthoryear{{Dubinski}}{{Dubinski}}{1998}]{1998ApJ...502..141D}
{Dubinski} J.,  1998, \mn@doi [\apj] {10.1086/305901}, \href
  {http://adsabs.harvard.edu/abs/1998ApJ...502..141D} {502, 141}

\bibitem[\protect\citeauthoryear{{Dubois}, {Devriendt}, {Slyz}  \&
  {Teyssier}}{{Dubois} et~al.}{2010}]{2010MNRAS.409..985D}
{Dubois} Y.,  {Devriendt} J.,  {Slyz} A.,   {Teyssier} R.,  2010, \mn@doi
  [\mnras] {10.1111/j.1365-2966.2010.17338.x}, \href
  {http://adsabs.harvard.edu/abs/2010MNRAS.409..985D} {409, 985}

\bibitem[\protect\citeauthoryear{{Dubois}, {Devriendt}, {Slyz}  \&
  {Teyssier}}{{Dubois} et~al.}{2012}]{2012MNRAS.420.2662D}
{Dubois} Y.,  {Devriendt} J.,  {Slyz} A.,   {Teyssier} R.,  2012, \mn@doi
  [\mnras] {10.1111/j.1365-2966.2011.20236.x}, \href
  {http://adsabs.harvard.edu/abs/2012MNRAS.420.2662D} {420, 2662}

\bibitem[\protect\citeauthoryear{{Dubois} et~al.,}{{Dubois}
  et~al.}{2014}]{dubois14}
{Dubois} Y.,  et~al., 2014, \mn@doi [\mnras] {10.1093/mnras/stu1227}, \href
  {http://adsabs.harvard.edu/abs/2014MNRAS.444.1453D} {444, 1453}

\bibitem[\protect\citeauthoryear{{Dubois}, {Peirani}, {Pichon}, {Devriendt},
  {Gavazzi}, {Welker}  \& {Volonteri}}{{Dubois}
  et~al.}{2016}]{2016MNRAS.463.3948D}
{Dubois} Y.,  {Peirani} S.,  {Pichon} C.,  {Devriendt} J.,  {Gavazzi} R.,
  {Welker} C.,   {Volonteri} M.,  2016, \mn@doi [\mnras]
  {10.1093/mnras/stw2265}, \href {http://ads.nao.ac.jp/abs/2016MNRAS.463.3948D}
  {463, 3948}

\bibitem[\protect\citeauthoryear{{Evans} \& {Bridle}}{{Evans} \&
  {Bridle}}{2009}]{2009ApJ...695.1446E}
{Evans} A.~K.~D.,  {Bridle} S.,  2009, \mn@doi [\apj]
  {10.1088/0004-637X/695/2/1446}, \href
  {http://adsabs.harvard.edu/abs/2009ApJ...695.1446E} {695, 1446}

\bibitem[\protect\citeauthoryear{{Faltenbacher}, {Li}, {Mao}, {van den Bosch},
  {Yang}, {Jing}, {Pasquali}  \& {Mo}}{{Faltenbacher}
  et~al.}{2007}]{2007ApJ...662L..71F}
{Faltenbacher} A.,  {Li} C.,  {Mao} S.,  {van den Bosch} F.~C.,  {Yang} X.,
  {Jing} Y.~P.,  {Pasquali} A.,   {Mo} H.~J.,  2007, \mn@doi [\apjl]
  {10.1086/519683}, \href {http://adsabs.harvard.edu/abs/2007ApJ...662L..71F}
  {662, L71}

\bibitem[\protect\citeauthoryear{{Faltenbacher}, {Jing}, {Li}, {Mao}, {Mo},
  {Pasquali}  \& {van den Bosch}}{{Faltenbacher}
  et~al.}{2008}]{2008ApJ...675..146F}
{Faltenbacher} A.,  {Jing} Y.~P.,  {Li} C.,  {Mao} S.,  {Mo} H.~J.,  {Pasquali}
  A.,   {van den Bosch} F.~C.,  2008, \mn@doi [\apj] {10.1086/525243}, \href
  {http://adsabs.harvard.edu/abs/2008ApJ...675..146F} {675, 146}

\bibitem[\protect\citeauthoryear{{Fasano}, {Pisani}, {Vio}  \&
  {Girardi}}{{Fasano} et~al.}{1993}]{1993ApJ...416..546F}
{Fasano} G.,  {Pisani} A.,  {Vio} R.,   {Girardi} M.,  1993, \mn@doi [\apj]
  {10.1086/173256}, \href {http://adsabs.harvard.edu/abs/1993ApJ...416..546F}
  {416, 546}

\bibitem[\protect\citeauthoryear{{Feng}}{{Feng}}{2010}]{2010ARA&A..48..495F}
{Feng} J.~L.,  2010, \mn@doi [\araa] {10.1146/annurev-astro-082708-101659},
  \href {http://adsabs.harvard.edu/abs/2010ARA%26A..48..495F} {48, 495}

\bibitem[\protect\citeauthoryear{{Flores}, {Allgood}, {Kravtsov}, {Primack},
  {Buote}  \& {Bullock}}{{Flores} et~al.}{2007}]{2007MNRAS.377..883F}
{Flores} R.~A.,  {Allgood} B.,  {Kravtsov} A.~V.,  {Primack} J.~R.,  {Buote}
  D.~A.,   {Bullock} J.~S.,  2007, \mn@doi [\mnras]
  {10.1111/j.1365-2966.2007.11658.x}, \href
  {http://adsabs.harvard.edu/abs/2007MNRAS.377..883F} {377, 883}

\bibitem[\protect\citeauthoryear{{Fo{\"e}x}, {Chon}  \&
  {B{\"o}hringer}}{{Fo{\"e}x} et~al.}{2017}]{2017A&A...601A.145F}
{Fo{\"e}x} G.,  {Chon} G.,   {B{\"o}hringer} H.,  2017, \mn@doi [\aap]
  {10.1051/0004-6361/201630086}, \href
  {http://adsabs.harvard.edu/abs/2017A%26A...601A.145F} {601, A145}

\bibitem[\protect\citeauthoryear{{Gottl{\"o}ber} \& {Yepes}}{{Gottl{\"o}ber} \&
  {Yepes}}{2007}]{2007ApJ...664..117G}
{Gottl{\"o}ber} S.,  {Yepes} G.,  2007, \mn@doi [\apj] {10.1086/517907}, \href
  {http://adsabs.harvard.edu/abs/2007ApJ...664..117G} {664, 117}

\bibitem[\protect\citeauthoryear{{Greggio} \& {Renzini}}{{Greggio} \&
  {Renzini}}{1983}]{1983A&A...118..217G}
{Greggio} L.,  {Renzini} A.,  1983, \aap, \href
  {http://adsabs.harvard.edu/abs/1983A%26A...118..217G} {118, 217}

\bibitem[\protect\citeauthoryear{{Hamana}, {Sakurai}, {Koike}  \&
  {Miller}}{{Hamana} et~al.}{2015}]{2015PASJ...67...34H}
{Hamana} T.,  {Sakurai} J.,  {Koike} M.,   {Miller} L.,  2015, \mn@doi [\pasj]
  {10.1093/pasj/psv034}, \href
  {http://adsabs.harvard.edu/abs/2015PASJ...67...34H} {67, 34}

\bibitem[\protect\citeauthoryear{{Hashimoto}, {Henry}  \&
  {Boehringer}}{{Hashimoto} et~al.}{2008}]{2008MNRAS.390.1562H}
{Hashimoto} Y.,  {Henry} J.~P.,   {Boehringer} H.,  2008, \mn@doi [\mnras]
  {10.1111/j.1365-2966.2008.13840.x}, \href
  {http://adsabs.harvard.edu/abs/2008MNRAS.390.1562H} {390, 1562}

\bibitem[\protect\citeauthoryear{{Hellwing}, {Cautun}, {Knebe}, {Juszkiewicz}
  \& {Knollmann}}{{Hellwing} et~al.}{2013}]{2013JCAP...10..012H}
{Hellwing} W.~A.,  {Cautun} M.,  {Knebe} A.,  {Juszkiewicz} R.,   {Knollmann}
  S.,  2013, \mn@doi [\jcap] {10.1088/1475-7516/2013/10/012}, \href
  {http://adsabs.harvard.edu/abs/2013JCAP...10..012H} {10, 012}

\bibitem[\protect\citeauthoryear{{Hildebrandt} et~al.,}{{Hildebrandt}
  et~al.}{2017}]{2017MNRAS.465.1454H}
{Hildebrandt} H.,  et~al., 2017, \mn@doi [\mnras] {10.1093/mnras/stw2805},
  \href {http://adsabs.harvard.edu/abs/2017MNRAS.465.1454H} {465, 1454}

\bibitem[\protect\citeauthoryear{{Hirata} \& {Seljak}}{{Hirata} \&
  {Seljak}}{2003}]{2003MNRAS.343..459H}
{Hirata} C.,  {Seljak} U.,  2003, \mn@doi [\mnras]
  {10.1046/j.1365-8711.2003.06683.x}, \href
  {http://adsabs.harvard.edu/abs/2003MNRAS.343..459H} {343, 459}

\bibitem[\protect\citeauthoryear{{Ho}, {Bahcall}  \& {Bode}}{{Ho}
  et~al.}{2006}]{2006ApJ...647....8H}
{Ho} S.,  {Bahcall} N.,   {Bode} P.,  2006, \mn@doi [\apj] {10.1086/505255},
  \href {http://adsabs.harvard.edu/abs/2006ApJ...647....8H} {647, 8}

\bibitem[\protect\citeauthoryear{{Horowitz} \& {Seljak}}{{Horowitz} \&
  {Seljak}}{2017}]{2017MNRAS.469..394H}
{Horowitz} B.,  {Seljak} U.,  2017, \mn@doi [\mnras] {10.1093/mnras/stx766},
  \href {http://adsabs.harvard.edu/abs/2017MNRAS.469..394H} {469, 394}

\bibitem[\protect\citeauthoryear{{Huang}, {Mandelbaum}, {Freeman}, {Chen},
  {Rozo}, {Rykoff}  \& {Baxter}}{{Huang} et~al.}{2016}]{2016MNRAS.463..222H}
{Huang} H.-J.,  {Mandelbaum} R.,  {Freeman} P.~E.,  {Chen} Y.-C.,  {Rozo} E.,
  {Rykoff} E.,   {Baxter} E.~J.,  2016, \mn@doi [\mnras]
  {10.1093/mnras/stw1982}, \href
  {http://adsabs.harvard.edu/abs/2016MNRAS.463..222H} {463, 222}

\bibitem[\protect\citeauthoryear{{Huff}, {Hirata}, {Mandelbaum}, {Schlegel},
  {Seljak}  \& {Lupton}}{{Huff} et~al.}{2014}]{2014MNRAS.440.1296H}
{Huff} E.~M.,  {Hirata} C.~M.,  {Mandelbaum} R.,  {Schlegel} D.,  {Seljak} U.,
   {Lupton} R.~H.,  2014, \mn@doi [\mnras] {10.1093/mnras/stu144}, \href
  {http://adsabs.harvard.edu/abs/2014MNRAS.440.1296H} {440, 1296}

\bibitem[\protect\citeauthoryear{{Jedrzejewski}}{{Jedrzejewski}}{1987}]{1987MNRAS.226..747J}
{Jedrzejewski} R.~I.,  1987, \mn@doi [\mnras] {10.1093/mnras/226.4.747}, \href
  {http://adsabs.harvard.edu/abs/1987MNRAS.226..747J} {226, 747}

\bibitem[\protect\citeauthoryear{{Jullo}, {Kneib}, {Limousin},
  {El{\'{\i}}asd{\'o}ttir}, {Marshall}  \& {Verdugo}}{{Jullo}
  et~al.}{2007}]{2007NJPh....9..447J}
{Jullo} E.,  {Kneib} J.-P.,  {Limousin} M.,  {El{\'{\i}}asd{\'o}ttir} {\'A}.,
  {Marshall} P.~J.,   {Verdugo} T.,  2007, \mn@doi [New Journal of Physics]
  {10.1088/1367-2630/9/12/447}, \href
  {http://adsabs.harvard.edu/abs/2007NJPh....9..447J} {9, 447}

\bibitem[\protect\citeauthoryear{{Kaastra}, {Mewe}  \&
  {Nieuwenhuijzen}}{{Kaastra} et~al.}{1996}]{1996uxsa.conf..411K}
{Kaastra} J.~S.,  {Mewe} R.,   {Nieuwenhuijzen} H.,  1996, in {Yamashita} K.,
  {Watanabe} T.,  eds, UV and X-ray Spectroscopy of Astrophysical and
  Laboratory Plasmas. pp 411--414

\bibitem[\protect\citeauthoryear{{Kaiser}, {Squires}  \& {Broadhurst}}{{Kaiser}
  et~al.}{1995}]{1995ApJ...449..460K}
{Kaiser} N.,  {Squires} G.,   {Broadhurst} T.,  1995, \mn@doi [\apj]
  {10.1086/176071}, \href {http://adsabs.harvard.edu/abs/1995ApJ...449..460K}
  {449, 460}

\bibitem[\protect\citeauthoryear{{Kang}, {van den Bosch}, {Yang}, {Mao}, {Mo},
  {Li}  \& {Jing}}{{Kang} et~al.}{2007}]{2007MNRAS.378.1531K}
{Kang} X.,  {van den Bosch} F.~C.,  {Yang} X.,  {Mao} S.,  {Mo} H.~J.,  {Li}
  C.,   {Jing} Y.~P.,  2007, \mn@doi [\mnras]
  {10.1111/j.1365-2966.2007.11902.x}, \href
  {http://adsabs.harvard.edu/abs/2007MNRAS.378.1531K} {378, 1531}

\bibitem[\protect\citeauthoryear{{Kasun} \& {Evrard}}{{Kasun} \&
  {Evrard}}{2005}]{2005ApJ...629..781K}
{Kasun} S.~F.,  {Evrard} A.~E.,  2005, \mn@doi [\apj] {10.1086/430811}, \href
  {http://adsabs.harvard.edu/abs/2005ApJ...629..781K} {629, 781}

\bibitem[\protect\citeauthoryear{{Kaviraj} et~al.,}{{Kaviraj}
  et~al.}{2017}]{2017MNRAS.467.4739K}
{Kaviraj} S.,  et~al., 2017, \mn@doi [\mnras] {10.1093/mnras/stx126}, \href
  {http://ads.nao.ac.jp/abs/2017MNRAS.467.4739K} {467, 4739}

\bibitem[\protect\citeauthoryear{{Kawahara}}{{Kawahara}}{2010}]{2010ApJ...719.1926K}
{Kawahara} H.,  2010, \mn@doi [\apj] {10.1088/0004-637X/719/2/1926}, \href
  {http://adsabs.harvard.edu/abs/2010ApJ...719.1926K} {719, 1926}

\bibitem[\protect\citeauthoryear{{Khandai}, {Di Matteo}, {Croft}, {Wilkins},
  {Feng}, {Tucker}, {DeGraf}  \& {Liu}}{{Khandai}
  et~al.}{2015}]{2015MNRAS.450.1349K}
{Khandai} N.,  {Di Matteo} T.,  {Croft} R.,  {Wilkins} S.,  {Feng} Y.,
  {Tucker} E.,  {DeGraf} C.,   {Liu} M.-S.,  2015, \mn@doi [\mnras]
  {10.1093/mnras/stv627}, \href
  {http://adsabs.harvard.edu/abs/2015MNRAS.450.1349K} {450, 1349}

\bibitem[\protect\citeauthoryear{{Khoury}}{{Khoury}}{2015}]{2015PhRvD..91b4022K}
{Khoury} J.,  2015, \mn@doi [\prd] {10.1103/PhysRevD.91.024022}, \href
  {http://adsabs.harvard.edu/abs/2015PhRvD..91b4022K} {91, 024022}

\bibitem[\protect\citeauthoryear{{Kneib}}{{Kneib}}{1993}]{1993PhDT.......189K}
{Kneib} J.-P.,  1993, PhD thesis, Ph.~D.~thesis, Universit{\'e} Paul Sabatier,
  Toulouse, (1993)

\bibitem[\protect\citeauthoryear{{Koester} et~al.,}{{Koester}
  et~al.}{2007}]{2007ApJ...660..239K}
{Koester} B.~P.,  et~al., 2007, \mn@doi [\apj] {10.1086/509599}, \href
  {http://adsabs.harvard.edu/abs/2007ApJ...660..239K} {660, 239}

\bibitem[\protect\citeauthoryear{{Komatsu} et~al.,}{{Komatsu}
  et~al.}{2011}]{WMAP}
{Komatsu} E.,  et~al., 2011, \mn@doi [\apjs] {10.1088/0067-0049/192/2/18},
  \href {http://adsabs.harvard.edu/abs/2011ApJS..192...18K} {192, 18}

\bibitem[\protect\citeauthoryear{{L'Huillier}, {Park}  \& {Kim}}{{L'Huillier}
  et~al.}{2017a}]{2017MNRAS.466.4875L}
{L'Huillier} B.,  {Park} C.,   {Kim} J.,  2017a, \mn@doi [\mnras]
  {10.1093/mnras/stx124}, \href
  {http://adsabs.harvard.edu/abs/2017MNRAS.466.4875L} {466, 4875}

\bibitem[\protect\citeauthoryear{{L'Huillier}, {Winther}, {Mota}, {Park}  \&
  {Kim}}{{L'Huillier} et~al.}{2017b}]{2017MNRAS.468.3174L}
{L'Huillier} B.,  {Winther} H.~A.,  {Mota} D.~F.,  {Park} C.,   {Kim} J.,
  2017b, \mn@doi [\mnras] {10.1093/mnras/stx700}, \href
  {http://adsabs.harvard.edu/abs/2017MNRAS.468.3174L} {468, 3174}

\bibitem[\protect\citeauthoryear{{LaRoque}, {Bonamente}, {Carlstrom}, {Joy},
  {Nagai}, {Reese}  \& {Dawson}}{{LaRoque} et~al.}{2006}]{2006ApJ...652..917L}
{LaRoque} S.~J.,  {Bonamente} M.,  {Carlstrom} J.~E.,  {Joy} M.~K.,  {Nagai}
  D.,  {Reese} E.~D.,   {Dawson} K.~S.,  2006, \mn@doi [\apj] {10.1086/508139},
  \href {http://adsabs.harvard.edu/abs/2006ApJ...652..917L} {652, 917}

\bibitem[\protect\citeauthoryear{{Lambas}, {Groth}  \& {Peebles}}{{Lambas}
  et~al.}{1988}]{1988AJ.....95..996L}
{Lambas} D.~G.,  {Groth} E.~J.,   {Peebles} P.~J.~E.,  1988, \mn@doi [\aj]
  {10.1086/114695}, \href {http://adsabs.harvard.edu/abs/1988AJ.....95..996L}
  {95, 996}

\bibitem[\protect\citeauthoryear{{Lau}, {Nagai}, {Kravtsov}, {Vikhlinin}  \&
  {Zentner}}{{Lau} et~al.}{2012}]{2012ApJ...755..116L}
{Lau} E.~T.,  {Nagai} D.,  {Kravtsov} A.~V.,  {Vikhlinin} A.,   {Zentner}
  A.~R.,  2012, \mn@doi [\apj] {10.1088/0004-637X/755/2/116}, \href
  {http://adsabs.harvard.edu/abs/2012ApJ...755..116L} {755, 116}

\bibitem[\protect\citeauthoryear{{Le Brun}, {McCarthy}, {Schaye}  \&
  {Ponman}}{{Le Brun} et~al.}{2014}]{2014MNRAS.441.1270L}
{Le Brun} A.~M.~C.,  {McCarthy} I.~G.,  {Schaye} J.,   {Ponman} T.~J.,  2014,
  \mn@doi [\mnras] {10.1093/mnras/stu608}, \href
  {http://adsabs.harvard.edu/abs/2014MNRAS.441.1270L} {441, 1270}

\bibitem[\protect\citeauthoryear{{Leitherer} et~al.,}{{Leitherer}
  et~al.}{1999}]{1999ApJS..123....3L}
{Leitherer} C.,  et~al., 1999, \mn@doi [\apjs] {10.1086/313233}, \href
  {http://adsabs.harvard.edu/abs/1999ApJS..123....3L} {123, 3}

\bibitem[\protect\citeauthoryear{{Leitherer}, {Ortiz Ot{\'a}lvaro}, {Bresolin},
  {Kudritzki}, {Lo Faro}, {Pauldrach}, {Pettini}  \& {Rix}}{{Leitherer}
  et~al.}{2010}]{2010ApJS..189..309L}
{Leitherer} C.,  {Ortiz Ot{\'a}lvaro} P.~A.,  {Bresolin} F.,  {Kudritzki}
  R.-P.,  {Lo Faro} B.,  {Pauldrach} A.~W.~A.,  {Pettini} M.,   {Rix} S.~A.,
  2010, \mn@doi [\apjs] {10.1088/0067-0049/189/2/309}, \href
  {http://adsabs.harvard.edu/abs/2010ApJS..189..309L} {189, 309}

\bibitem[\protect\citeauthoryear{{Li}, {Wang}, {Yang}, {Chen}, {Xie}  \&
  {Wang}}{{Li} et~al.}{2013}]{2013ApJ...768...20L}
{Li} Z.,  {Wang} Y.,  {Yang} X.,  {Chen} X.,  {Xie} L.,   {Wang} X.,  2013,
  \mn@doi [\apj] {10.1088/0004-637X/768/1/20}, \href
  {http://adsabs.harvard.edu/abs/2013ApJ...768...20L} {768, 20}

\bibitem[\protect\citeauthoryear{{Liu} et~al.,}{{Liu}
  et~al.}{2015}]{2015MNRAS.450.2888L}
{Liu} X.,  et~al., 2015, \mn@doi [\mnras] {10.1093/mnras/stv784}, \href
  {http://adsabs.harvard.edu/abs/2015MNRAS.450.2888L} {450, 2888}

\bibitem[\protect\citeauthoryear{{Lovisari} et~al.,}{{Lovisari}
  et~al.}{2017}]{2017ApJ...846...51L}
{Lovisari} L.,  et~al., 2017, \mn@doi [\apj] {10.3847/1538-4357/aa855f}, \href
  {http://adsabs.harvard.edu/abs/2017ApJ...846...51L} {846, 51}

\bibitem[\protect\citeauthoryear{{Mandelbaum}, {Slosar}, {Baldauf}, {Seljak},
  {Hirata}, {Nakajima}, {Reyes}  \& {Smith}}{{Mandelbaum}
  et~al.}{2013}]{2013MNRAS.432.1544M}
{Mandelbaum} R.,  {Slosar} A.,  {Baldauf} T.,  {Seljak} U.,  {Hirata} C.~M.,
  {Nakajima} R.,  {Reyes} R.,   {Smith} R.~E.,  2013, \mn@doi [\mnras]
  {10.1093/mnras/stt572}, \href
  {http://adsabs.harvard.edu/abs/2013MNRAS.432.1544M} {432, 1544}

\bibitem[\protect\citeauthoryear{{Mantz}, {Allen}, {Morris}, {Rapetti},
  {Applegate}, {Kelly}, {von der Linden}  \& {Schmidt}}{{Mantz}
  et~al.}{2014}]{2014MNRAS.440.2077M}
{Mantz} A.~B.,  {Allen} S.~W.,  {Morris} R.~G.,  {Rapetti} D.~A.,  {Applegate}
  D.~E.,  {Kelly} P.~L.,  {von der Linden} A.,   {Schmidt} R.~W.,  2014,
  \mn@doi [\mnras] {10.1093/mnras/stu368}, \href
  {http://adsabs.harvard.edu/abs/2014MNRAS.440.2077M} {440, 2077}

\bibitem[\protect\citeauthoryear{{Monaghan}}{{Monaghan}}{1992}]{1992ARA&A..30..543M}
{Monaghan} J.~J.,  1992, \mn@doi [\araa] {10.1146/annurev.aa.30.090192.002551},
  \href {http://adsabs.harvard.edu/abs/1992ARA%26A..30..543M} {30, 543}

\bibitem[\protect\citeauthoryear{{Navarro}, {Frenk}  \& {White}}{{Navarro}
  et~al.}{1997}]{1997ApJ...490..493N}
{Navarro} J.~F.,  {Frenk} C.~S.,   {White} S.~D.~M.,  1997, \mn@doi [\apj]
  {10.1086/304888}, \href {http://adsabs.harvard.edu/abs/1997ApJ...490..493N}
  {490, 493}

\bibitem[\protect\citeauthoryear{{Oguri}}{{Oguri}}{2010}]{2010PASJ...62.1017O}
{Oguri} M.,  2010, \mn@doi [\pasj] {10.1093/pasj/62.4.1017}, \href
  {http://adsabs.harvard.edu/abs/2010PASJ...62.1017O} {62, 1017}

\bibitem[\protect\citeauthoryear{{Oguri} et~al.,}{{Oguri}
  et~al.}{2009}]{2009ApJ...699.1038O}
{Oguri} M.,  et~al., 2009, \mn@doi [\apj] {10.1088/0004-637X/699/2/1038}, \href
  {http://adsabs.harvard.edu/abs/2009ApJ...699.1038O} {699, 1038}

\bibitem[\protect\citeauthoryear{{Oguri}, {Takada}, {Okabe}  \&
  {Smith}}{{Oguri} et~al.}{2010}]{2010MNRAS.405.2215O}
{Oguri} M.,  {Takada} M.,  {Okabe} N.,   {Smith} G.~P.,  2010, \mn@doi [\mnras]
  {10.1111/j.1365-2966.2010.16622.x}, \href
  {http://adsabs.harvard.edu/abs/2010MNRAS.405.2215O} {405, 2215}

\bibitem[\protect\citeauthoryear{{Oguri}, {Bayliss}, {Dahle}, {Sharon},
  {Gladders}, {Natarajan}, {Hennawi}  \& {Koester}}{{Oguri}
  et~al.}{2012}]{2012MNRAS.420.3213O}
{Oguri} M.,  {Bayliss} M.~B.,  {Dahle} H.,  {Sharon} K.,  {Gladders} M.~D.,
  {Natarajan} P.,  {Hennawi} J.~F.,   {Koester} B.~P.,  2012, \mn@doi [\mnras]
  {10.1111/j.1365-2966.2011.20248.x}, \href
  {http://adsabs.harvard.edu/abs/2012MNRAS.420.3213O} {420, 3213}

\bibitem[\protect\citeauthoryear{{Okabe}, {Takada}, {Umetsu}, {Futamase}  \&
  {Smith}}{{Okabe} et~al.}{2010}]{2010PASJ...62..811O}
{Okabe} N.,  {Takada} M.,  {Umetsu} K.,  {Futamase} T.,   {Smith} G.~P.,  2010,
  \mn@doi [\pasj] {10.1093/pasj/62.3.811}, \href
  {http://adsabs.harvard.edu/abs/2010PASJ...62..811O} {62, 811}

\bibitem[\protect\citeauthoryear{{Omma}, {Binney}, {Bryan}  \& {Slyz}}{{Omma}
  et~al.}{2004}]{2004MNRAS.348.1105O}
{Omma} H.,  {Binney} J.,  {Bryan} G.,   {Slyz} A.,  2004, \mn@doi [\mnras]
  {10.1111/j.1365-2966.2004.07382.x}, \href
  {http://adsabs.harvard.edu/abs/2004MNRAS.348.1105O} {348, 1105}

\bibitem[\protect\citeauthoryear{{Osato}, {Nishimichi}, {Oguri}, {Takada}  \&
  {Okumura}}{{Osato} et~al.}{2017}]{2017arXiv171200094O}
{Osato} K.,  {Nishimichi} T.,  {Oguri} M.,  {Takada} M.,   {Okumura} T.,  2017,
  preprint, \href {http://adsabs.harvard.edu/abs/2017arXiv171200094O} {}
  (\mn@eprint {arXiv} {1712.00094})

\bibitem[\protect\citeauthoryear{{Panko}, {Juszczyk}, {Biernacka}  \&
  {Flin}}{{Panko} et~al.}{2009}]{2009ApJ...700.1686P}
{Panko} E.,  {Juszczyk} T.,  {Biernacka} M.,   {Flin} P.,  2009, \mn@doi [\apj]
  {10.1088/0004-637X/700/2/1686}, \href
  {http://adsabs.harvard.edu/abs/2009ApJ...700.1686P} {700, 1686}

\bibitem[\protect\citeauthoryear{{Parekh}, {van der Heyden}, {Ferrari}, {Angus}
   \& {Holwerda}}{{Parekh} et~al.}{2015}]{2015A&A...575A.127P}
{Parekh} V.,  {van der Heyden} K.,  {Ferrari} C.,  {Angus} G.,   {Holwerda} B.,
   2015, \mn@doi [\aap] {10.1051/0004-6361/201424123}, \href
  {http://adsabs.harvard.edu/abs/2015A%26A...575A.127P} {575, A127}

\bibitem[\protect\citeauthoryear{{Paz}, {Lambas}, {Padilla}  \&
  {Merch{\'a}n}}{{Paz} et~al.}{2006}]{2006MNRAS.366.1503P}
{Paz} D.~J.,  {Lambas} D.~G.,  {Padilla} N.,   {Merch{\'a}n} M.,  2006, \mn@doi
  [\mnras] {10.1111/j.1365-2966.2005.09934.x}, \href
  {http://adsabs.harvard.edu/abs/2006MNRAS.366.1503P} {366, 1503}

\bibitem[\protect\citeauthoryear{{Peirani} et~al.,}{{Peirani}
  et~al.}{2017}]{2017MNRAS.472.2153P}
{Peirani} S.,  et~al., 2017, \mn@doi [\mnras] {10.1093/mnras/stx2099}, \href
  {http://ads.nao.ac.jp/abs/2017MNRAS.472.2153P} {472, 2153}

\bibitem[\protect\citeauthoryear{{Peirani} et~al.,}{{Peirani}
  et~al.}{2018}]{2018arXiv180109754P}
{Peirani} S.,  et~al., 2018, preprint, \href
  {http://adsabs.harvard.edu/abs/2018arXiv180109754P} {} (\mn@eprint {arXiv}
  {1801.09754})

\bibitem[\protect\citeauthoryear{{Peter}, {Rocha}, {Bullock}  \&
  {Kaplinghat}}{{Peter} et~al.}{2013}]{2013MNRAS.430..105P}
{Peter} A.~H.~G.,  {Rocha} M.,  {Bullock} J.~S.,   {Kaplinghat} M.,  2013,
  \mn@doi [\mnras] {10.1093/mnras/sts535}, \href
  {http://adsabs.harvard.edu/abs/2013MNRAS.430..105P} {430, 105}

\bibitem[\protect\citeauthoryear{{Piras}, {Joachimi}, {Sch{\"a}fer},
  {Bonamigo}, {Hilbert}  \& {van Uitert}}{{Piras}
  et~al.}{2018}]{2018MNRAS.474.1165P}
{Piras} D.,  {Joachimi} B.,  {Sch{\"a}fer} B.~M.,  {Bonamigo} M.,  {Hilbert}
  S.,   {van Uitert} E.,  2018, \mn@doi [\mnras] {10.1093/mnras/stx2846}, \href
  {http://adsabs.harvard.edu/abs/2018MNRAS.474.1165P} {474, 1165}

\bibitem[\protect\citeauthoryear{{Planck Collaboration} et~al.,}{{Planck
  Collaboration} et~al.}{2014}]{2014A&A...571A..20P}
{Planck Collaboration} et~al., 2014, \mn@doi [\aap]
  {10.1051/0004-6361/201321521}, \href
  {http://adsabs.harvard.edu/abs/2014A%26A...571A..20P} {571, A20}

\bibitem[\protect\citeauthoryear{{Planck Collaboration} et~al.,}{{Planck
  Collaboration} et~al.}{2016}]{2016A&A...594A..24P}
{Planck Collaboration} et~al., 2016, \mn@doi [\aap]
  {10.1051/0004-6361/201525833}, \href
  {http://adsabs.harvard.edu/abs/2016A%26A...594A..24P} {594, A24}

\bibitem[\protect\citeauthoryear{{Porter}, {Schneider}  \& {Hoessel}}{{Porter}
  et~al.}{1991}]{1991AJ....101.1561P}
{Porter} A.~C.,  {Schneider} D.~P.,   {Hoessel} J.~G.,  1991, \mn@doi [\aj]
  {10.1086/115788}, \href {http://adsabs.harvard.edu/abs/1991AJ....101.1561P}
  {101, 1561}

\bibitem[\protect\citeauthoryear{{Prunet}, {Pichon}, {Aubert}, {Pogosyan},
  {Teyssier}  \& {Gottloeber}}{{Prunet} et~al.}{2008}]{2008ApJS..178..179P}
{Prunet} S.,  {Pichon} C.,  {Aubert} D.,  {Pogosyan} D.,  {Teyssier} R.,
  {Gottloeber} S.,  2008, \mn@doi [\apjs] {10.1086/590370}, \href
  {http://adsabs.harvard.edu/abs/2008ApJS..178..179P} {178, 179}

\bibitem[\protect\citeauthoryear{{Rahman}, {Shandarin}, {Motl}  \&
  {Melott}}{{Rahman} et~al.}{2004}]{2004astro.ph..5097R}
{Rahman} N.,  {Shandarin} S.~F.,  {Motl} P.~M.,   {Melott} A.~L.,  2004, ArXiv
  Astrophysics e-prints, \href
  {http://adsabs.harvard.edu/abs/2004astro.ph..5097R} {}

\bibitem[\protect\citeauthoryear{{Reichardt} et~al.,}{{Reichardt}
  et~al.}{2013}]{2013ApJ...763..127R}
{Reichardt} C.~L.,  et~al., 2013, \mn@doi [\apj] {10.1088/0004-637X/763/2/127},
  \href {http://adsabs.harvard.edu/abs/2013ApJ...763..127R} {763, 127}

\bibitem[\protect\citeauthoryear{{Reyes}, {Mandelbaum}, {Gunn}, {Nakajima},
  {Seljak}  \& {Hirata}}{{Reyes} et~al.}{2012}]{2012MNRAS.425.2610R}
{Reyes} R.,  {Mandelbaum} R.,  {Gunn} J.~E.,  {Nakajima} R.,  {Seljak} U.,
  {Hirata} C.~M.,  2012, \mn@doi [\mnras] {10.1111/j.1365-2966.2012.21472.x},
  \href {http://adsabs.harvard.edu/abs/2012MNRAS.425.2610R} {425, 2610}

\bibitem[\protect\citeauthoryear{{Rhee} \& {Katgert}}{{Rhee} \&
  {Katgert}}{1987}]{1987A&A...183..217R}
{Rhee} G.~F.~R.~N.,  {Katgert} P.,  1987, \aap, \href
  {http://adsabs.harvard.edu/abs/1987A%26A...183..217R} {183, 217}

\bibitem[\protect\citeauthoryear{{Rhee} \& {Latour}}{{Rhee} \&
  {Latour}}{1991}]{1991A&A...243...38R}
{Rhee} G.~F.~R.~N.,  {Latour} H.~J.,  1991, \aap, \href
  {http://adsabs.harvard.edu/abs/1991A%26A...243...38R} {243, 38}

\bibitem[\protect\citeauthoryear{{Richard} et~al.,}{{Richard}
  et~al.}{2010}]{2010MNRAS.404..325R}
{Richard} J.,  et~al., 2010, \mn@doi [\mnras]
  {10.1111/j.1365-2966.2009.16274.x}, \href
  {http://adsabs.harvard.edu/abs/2010MNRAS.404..325R} {404, 325}

\bibitem[\protect\citeauthoryear{{Rocha}, {Peter}, {Bullock}, {Kaplinghat},
  {Garrison-Kimmel}, {O{\~n}orbe}  \& {Moustakas}}{{Rocha}
  et~al.}{2013}]{2013MNRAS.430...81R}
{Rocha} M.,  {Peter} A.~H.~G.,  {Bullock} J.~S.,  {Kaplinghat} M.,
  {Garrison-Kimmel} S.,  {O{\~n}orbe} J.,   {Moustakas} L.~A.,  2013, \mn@doi
  [\mnras] {10.1093/mnras/sts514}, \href
  {http://adsabs.harvard.edu/abs/2013MNRAS.430...81R} {430, 81}

\bibitem[\protect\citeauthoryear{{Rozo} et~al.,}{{Rozo}
  et~al.}{2010}]{2010ApJ...708..645R}
{Rozo} E.,  et~al., 2010, \mn@doi [\apj] {10.1088/0004-637X/708/1/645}, \href
  {http://adsabs.harvard.edu/abs/2010ApJ...708..645R} {708, 645}

\bibitem[\protect\citeauthoryear{{Salpeter}}{{Salpeter}}{1955}]{1955ApJ...121..161S}
{Salpeter} E.~E.,  1955, \mn@doi [\apj] {10.1086/145971}, \href
  {http://adsabs.harvard.edu/abs/1955ApJ...121..161S} {121, 161}

\bibitem[\protect\citeauthoryear{{Sastry}}{{Sastry}}{1968}]{1968PASP...80..252S}
{Sastry} G.~N.,  1968, \mn@doi [\pasp] {10.1086/128626}, \href
  {http://adsabs.harvard.edu/abs/1968PASP...80..252S} {80, 252}

\bibitem[\protect\citeauthoryear{{Sayers} et~al.,}{{Sayers}
  et~al.}{2013}]{2013ApJ...768..177S}
{Sayers} J.,  et~al., 2013, \mn@doi [\apj] {10.1088/0004-637X/768/2/177}, \href
  {http://adsabs.harvard.edu/abs/2013ApJ...768..177S} {768, 177}

\bibitem[\protect\citeauthoryear{{Schaller} et~al.,}{{Schaller}
  et~al.}{2015}]{2015MNRAS.452..343S}
{Schaller} M.,  et~al., 2015, \mn@doi [\mnras] {10.1093/mnras/stv1341}, \href
  {http://adsabs.harvard.edu/abs/2015MNRAS.452..343S} {452, 343}

\bibitem[\protect\citeauthoryear{{Schaye} et~al.,}{{Schaye}
  et~al.}{2015}]{2015MNRAS.446..521S}
{Schaye} J.,  et~al., 2015, \mn@doi [\mnras] {10.1093/mnras/stu2058}, \href
  {http://adsabs.harvard.edu/abs/2015MNRAS.446..521S} {446, 521}

\bibitem[\protect\citeauthoryear{{Schellenberger} \&
  {Reiprich}}{{Schellenberger} \& {Reiprich}}{2017}]{2017MNRAS.469.3738S}
{Schellenberger} G.,  {Reiprich} T.~H.,  2017, \mn@doi [\mnras]
  {10.1093/mnras/stx1022}, \href
  {http://adsabs.harvard.edu/abs/2017MNRAS.469.3738S} {469, 3738}

\bibitem[\protect\citeauthoryear{{Schmidt}, {Allen}  \& {Fabian}}{{Schmidt}
  et~al.}{2004}]{2004MNRAS.352.1413S}
{Schmidt} R.~W.,  {Allen} S.~W.,   {Fabian} A.~C.,  2004, \mn@doi [\mnras]
  {10.1111/j.1365-2966.2004.08032.x}, \href
  {http://adsabs.harvard.edu/abs/2004MNRAS.352.1413S} {352, 1413}

\bibitem[\protect\citeauthoryear{{Schneider}, {Frenk}  \& {Cole}}{{Schneider}
  et~al.}{2012}]{2012JCAP...05..030S}
{Schneider} M.~D.,  {Frenk} C.~S.,   {Cole} S.,  2012, \mn@doi [\jcap]
  {10.1088/1475-7516/2012/05/030}, \href
  {http://adsabs.harvard.edu/abs/2012JCAP...05..030S} {5, 030}

\bibitem[\protect\citeauthoryear{{Sereno}, {Umetsu}, {Ettori}, {Sayers},
  {Chiu}, {Meneghetti}, {Vega-Ferrero}  \& {Zitrin}}{{Sereno}
  et~al.}{2018}]{2018arXiv180400667S}
{Sereno} M.,  {Umetsu} K.,  {Ettori} S.,  {Sayers} J.,  {Chiu} I.,
  {Meneghetti} M.,  {Vega-Ferrero} J.,   {Zitrin} A.,  2018, preprint, \href
  {http://adsabs.harvard.edu/abs/2018arXiv180400667S} {} (\mn@eprint {arXiv}
  {1804.00667})

\bibitem[\protect\citeauthoryear{{Shakura} \& {Sunyaev}}{{Shakura} \&
  {Sunyaev}}{1973}]{1973A&A....24..337S}
{Shakura} N.~I.,  {Sunyaev} R.~A.,  1973, \aap, \href
  {http://adsabs.harvard.edu/abs/1973A%26A....24..337S} {24, 337}

\bibitem[\protect\citeauthoryear{{Shan} et~al.,}{{Shan}
  et~al.}{2018}]{2018MNRAS.474.1116S}
{Shan} H.,  et~al., 2018, \mn@doi [\mnras] {10.1093/mnras/stx2837}, \href
  {http://adsabs.harvard.edu/abs/2018MNRAS.474.1116S} {474, 1116}

\bibitem[\protect\citeauthoryear{{Sheldon}, {Cunha}, {Mandelbaum}, {Brinkmann}
  \& {Weaver}}{{Sheldon} et~al.}{2012}]{2012ApJS..201...32S}
{Sheldon} E.~S.,  {Cunha} C.~E.,  {Mandelbaum} R.,  {Brinkmann} J.,   {Weaver}
  B.~A.,  2012, \mn@doi [\apjs] {10.1088/0067-0049/201/2/32}, \href
  {http://adsabs.harvard.edu/abs/2012ApJS..201...32S} {201, 32}

\bibitem[\protect\citeauthoryear{{Shin}, {Clampitt}, {Jain}, {Bernstein},
  {Neil}, {Rozo}  \& {Rykoff}}{{Shin} et~al.}{2017}]{2017arXiv170511167S}
{Shin} T.-h.,  {Clampitt} J.,  {Jain} B.,  {Bernstein} G.,  {Neil} A.,  {Rozo}
  E.,   {Rykoff} E.,  2017, preprint, \href
  {http://adsabs.harvard.edu/abs/2017arXiv170511167S} {} (\mn@eprint {arXiv}
  {1705.11167})

\bibitem[\protect\citeauthoryear{{Shin}, {Clampitt}, {Jain}, {Bernstein},
  {Neil}, {Rozo}  \& {Rykoff}}{{Shin} et~al.}{2018}]{2018MNRAS.475.2421S}
{Shin} T.-h.,  {Clampitt} J.,  {Jain} B.,  {Bernstein} G.,  {Neil} A.,  {Rozo}
  E.,   {Rykoff} E.,  2018, \mn@doi [\mnras] {10.1093/mnras/stx3366}, \href
  {http://adsabs.harvard.edu/abs/2018MNRAS.475.2421S} {475, 2421}

\bibitem[\protect\citeauthoryear{{Simionescu} et~al.,}{{Simionescu}
  et~al.}{2011}]{2011Sci...331.1576S}
{Simionescu} A.,  et~al., 2011, \mn@doi [Science] {10.1126/science.1200331},
  \href {http://adsabs.harvard.edu/abs/2011Sci...331.1576S} {331, 1576}

\bibitem[\protect\citeauthoryear{{Snowden}, {Mushotzky}, {Kuntz}  \&
  {Davis}}{{Snowden} et~al.}{2008}]{2008A&A...478..615S}
{Snowden} S.~L.,  {Mushotzky} R.~F.,  {Kuntz} K.~D.,   {Davis} D.~S.,  2008,
  \mn@doi [\aap] {10.1051/0004-6361:20077930}, \href
  {http://adsabs.harvard.edu/abs/2008A%26A...478..615S} {478, 615}

\bibitem[\protect\citeauthoryear{{Song} \& {Lee}}{{Song} \&
  {Lee}}{2012}]{2012ApJ...748...98S}
{Song} H.,  {Lee} J.,  2012, \mn@doi [\apj] {10.1088/0004-637X/748/2/98}, \href
  {http://adsabs.harvard.edu/abs/2012ApJ...748...98S} {748, 98}

\bibitem[\protect\citeauthoryear{{Soucail}, {Fo{\"e}x}, {Pointecouteau},
  {Arnaud}  \& {Limousin}}{{Soucail} et~al.}{2015}]{2015A&A...581A..31S}
{Soucail} G.,  {Fo{\"e}x} G.,  {Pointecouteau} E.,  {Arnaud} M.,   {Limousin}
  M.,  2015, \mn@doi [\aap] {10.1051/0004-6361/201322689}, \href
  {http://adsabs.harvard.edu/abs/2015A%26A...581A..31S} {581, A31}

\bibitem[\protect\citeauthoryear{{Spergel} \& {Steinhardt}}{{Spergel} \&
  {Steinhardt}}{2000}]{2000PhRvL..84.3760S}
{Spergel} D.~N.,  {Steinhardt} P.~J.,  2000, \mn@doi [Physical Review Letters]
  {10.1103/PhysRevLett.84.3760}, \href
  {http://adsabs.harvard.edu/abs/2000PhRvL..84.3760S} {84, 3760}

\bibitem[\protect\citeauthoryear{{Strazzullo}, {Paolillo}, {Longo}, {Puddu},
  {Djorgovski}, {De Carvalho}  \& {Gal}}{{Strazzullo}
  et~al.}{2005}]{2005MNRAS.359..191S}
{Strazzullo} V.,  {Paolillo} M.,  {Longo} G.,  {Puddu} E.,  {Djorgovski} S.~G.,
   {De Carvalho} R.~R.,   {Gal} R.~R.,  2005, \mn@doi [\mnras]
  {10.1111/j.1365-2966.2005.08880.x}, \href
  {http://adsabs.harvard.edu/abs/2005MNRAS.359..191S} {359, 191}

\bibitem[\protect\citeauthoryear{{Suto}, {Kitayama}, {Nishimichi}, {Sasaki}  \&
  {Suto}}{{Suto} et~al.}{2016}]{2016PASJ...68...97S}
{Suto} D.,  {Kitayama} T.,  {Nishimichi} T.,  {Sasaki} S.,   {Suto} Y.,  2016,
  \mn@doi [\pasj] {10.1093/pasj/psw088}, \href
  {http://adsabs.harvard.edu/abs/2016PASJ...68...97S} {68, 97}

\bibitem[\protect\citeauthoryear{{Suto}, {Peirani}, {Dubois}, {Kitayama},
  {Nishimichi}, {Sasaki}  \& {Suto}}{{Suto} et~al.}{2017}]{suto17}
{Suto} D.,  {Peirani} S.,  {Dubois} Y.,  {Kitayama} T.,  {Nishimichi} T.,
  {Sasaki} S.,   {Suto} Y.,  2017, \mn@doi [\pasj] {10.1093/pasj/psw118}, \href
  {http://ads.nao.ac.jp/abs/2017PASJ...69...14S} {69, 14}

\bibitem[\protect\citeauthoryear{{Suwa}, {Habe}, {Yoshikawa}  \&
  {Okamoto}}{{Suwa} et~al.}{2003}]{2003ApJ...588....7S}
{Suwa} T.,  {Habe} A.,  {Yoshikawa} K.,   {Okamoto} T.,  2003, \mn@doi [\apj]
  {10.1086/368375}, \href {http://adsabs.harvard.edu/abs/2003ApJ...588....7S}
  {588, 7}

\bibitem[\protect\citeauthoryear{{Tenneti}, {Mandelbaum}, {Di Matteo},
  {Kiessling}  \& {Khandai}}{{Tenneti} et~al.}{2015}]{2015MNRAS.453..469T}
{Tenneti} A.,  {Mandelbaum} R.,  {Di Matteo} T.,  {Kiessling} A.,   {Khandai}
  N.,  2015, \mn@doi [\mnras] {10.1093/mnras/stv1625}, \href
  {http://adsabs.harvard.edu/abs/2015MNRAS.453..469T} {453, 469}

\bibitem[\protect\citeauthoryear{{Teyssier}}{{Teyssier}}{2002}]{2002A&A...385..337T}
{Teyssier} R.,  2002, \mn@doi [\aap] {10.1051/0004-6361:20011817}, \href
  {http://adsabs.harvard.edu/abs/2002A%26A...385..337T} {385, 337}

\bibitem[\protect\citeauthoryear{{Tulin} \& {Yu}}{{Tulin} \&
  {Yu}}{2017}]{2017arXiv170502358T}
{Tulin} S.,  {Yu} H.-B.,  2017, preprint, \href
  {http://adsabs.harvard.edu/abs/2017arXiv170502358T} {} (\mn@eprint {arXiv}
  {1705.02358})

\bibitem[\protect\citeauthoryear{{Tweed}, {Devriendt}, {Blaizot}, {Colombi}  \&
  {Slyz}}{{Tweed} et~al.}{2009}]{2009A&A...506..647T}
{Tweed} D.,  {Devriendt} J.,  {Blaizot} J.,  {Colombi} S.,   {Slyz} A.,  2009,
  \mn@doi [\aap] {10.1051/0004-6361/200911787}, \href
  {http://adsabs.harvard.edu/abs/2009A%26A...506..647T} {506, 647}

\bibitem[\protect\citeauthoryear{{Umetsu} et~al.,}{{Umetsu}
  et~al.}{2018}]{2018arXiv180400664U}
{Umetsu} K.,  et~al., 2018, preprint, \href
  {http://adsabs.harvard.edu/abs/2018arXiv180400664U} {} (\mn@eprint {arXiv}
  {1804.00664})

\bibitem[\protect\citeauthoryear{{Vega-Ferrero}, {Yepes}  \&
  {Gottl{\"o}ber}}{{Vega-Ferrero} et~al.}{2017}]{2017MNRAS.467.3226V}
{Vega-Ferrero} J.,  {Yepes} G.,   {Gottl{\"o}ber} S.,  2017, \mn@doi [\mnras]
  {10.1093/mnras/stx282}, \href
  {http://adsabs.harvard.edu/abs/2017MNRAS.467.3226V} {467, 3226}

\bibitem[\protect\citeauthoryear{{Velliscig} et~al.,}{{Velliscig}
  et~al.}{2015a}]{2015MNRAS.453..721V}
{Velliscig} M.,  et~al., 2015a, \mn@doi [\mnras] {10.1093/mnras/stv1690}, \href
  {http://adsabs.harvard.edu/abs/2015MNRAS.453..721V} {453, 721}

\bibitem[\protect\citeauthoryear{{Velliscig} et~al.,}{{Velliscig}
  et~al.}{2015b}]{2015MNRAS.454.3328V}
{Velliscig} M.,  et~al., 2015b, \mn@doi [\mnras] {10.1093/mnras/stv2198}, \href
  {http://adsabs.harvard.edu/abs/2015MNRAS.454.3328V} {454, 3328}

\bibitem[\protect\citeauthoryear{{Vikhlinin} et~al.,}{{Vikhlinin}
  et~al.}{2009}]{2009ApJ...692.1060V}
{Vikhlinin} A.,  et~al., 2009, \mn@doi [\apj] {10.1088/0004-637X/692/2/1060},
  \href {http://adsabs.harvard.edu/abs/2009ApJ...692.1060V} {692, 1060}

\bibitem[\protect\citeauthoryear{{Volonteri}, {Dubois}, {Pichon}  \&
  {Devriendt}}{{Volonteri} et~al.}{2016}]{2016MNRAS.460.2979V}
{Volonteri} M.,  {Dubois} Y.,  {Pichon} C.,   {Devriendt} J.,  2016, \mn@doi
  [\mnras] {10.1093/mnras/stw1123}, \href
  {http://ads.nao.ac.jp/abs/2016MNRAS.460.2979V} {460, 2979}

\bibitem[\protect\citeauthoryear{{Wei}, {Wu}  \& {Melia}}{{Wei}
  et~al.}{2015}]{2015MNRAS.447..479W}
{Wei} J.-J.,  {Wu} X.-F.,   {Melia} F.,  2015, \mn@doi [\mnras]
  {10.1093/mnras/stu2470}, \href
  {http://adsabs.harvard.edu/abs/2015MNRAS.447..479W} {447, 479}

\bibitem[\protect\citeauthoryear{{West}}{{West}}{1994}]{1994MNRAS.268...79W}
{West} M.~J.,  1994, \mn@doi [\mnras] {10.1093/mnras/268.1.79}, \href
  {http://adsabs.harvard.edu/abs/1994MNRAS.268...79W} {268, 79}

\bibitem[\protect\citeauthoryear{{West} \& {Blakeslee}}{{West} \&
  {Blakeslee}}{2000}]{2000ApJ...543L..27W}
{West} M.~J.,  {Blakeslee} J.~P.,  2000, \mn@doi [\apjl] {10.1086/318177},
  \href {http://adsabs.harvard.edu/abs/2000ApJ...543L..27W} {543, L27}

\bibitem[\protect\citeauthoryear{{West}, {Jones}  \& {Forman}}{{West}
  et~al.}{1995}]{1995ApJ...451L...5W}
{West} M.~J.,  {Jones} C.,   {Forman} W.,  1995, \mn@doi [\apjl]
  {10.1086/309673}, \href {http://adsabs.harvard.edu/abs/1995ApJ...451L...5W}
  {451, L5}

\bibitem[\protect\citeauthoryear{{West}, {de Propris}, {Bremer}  \&
  {Phillipps}}{{West} et~al.}{2017}]{2017NatAs...1E.157W}
{West} M.~J.,  {de Propris} R.,  {Bremer} M.~N.,   {Phillipps} S.,  2017,
  \mn@doi [Nature Astronomy] {10.1038/s41550-017-0157}, \href
  {http://adsabs.harvard.edu/abs/2017NatAs...1E.157W} {1, 0157}

\bibitem[\protect\citeauthoryear{{Yang}, {van den Bosch}, {Mo}, {Mao}, {Kang},
  {Weinmann}, {Guo}  \& {Jing}}{{Yang} et~al.}{2006}]{2006MNRAS.369.1293Y}
{Yang} X.,  {van den Bosch} F.~C.,  {Mo} H.~J.,  {Mao} S.,  {Kang} X.,
  {Weinmann} S.~M.,  {Guo} Y.,   {Jing} Y.~P.,  2006, \mn@doi [\mnras]
  {10.1111/j.1365-2966.2006.10373.x}, \href
  {http://adsabs.harvard.edu/abs/2006MNRAS.369.1293Y} {369, 1293}

\bibitem[\protect\citeauthoryear{{Yoshida}, {Springel}, {White}  \&
  {Tormen}}{{Yoshida} et~al.}{2000}]{2000ApJ...535L.103Y}
{Yoshida} N.,  {Springel} V.,  {White} S.~D.~M.,   {Tormen} G.,  2000, \mn@doi
  [\apjl] {10.1086/312707}, \href
  {http://adsabs.harvard.edu/abs/2000ApJ...535L.103Y} {535, L103}

\bibitem[\protect\citeauthoryear{{Zitrin} et~al.,}{{Zitrin}
  et~al.}{2015}]{2015ApJ...801...44Z}
{Zitrin} A.,  et~al., 2015, \mn@doi [\apj] {10.1088/0004-637X/801/1/44}, \href
  {http://adsabs.harvard.edu/abs/2015ApJ...801...44Z} {801, 44}

\bibitem[\protect\citeauthoryear{{Zu}, {Weinberg}, {Rozo}, {Sheldon}, {Tinker}
  \& {Becker}}{{Zu} et~al.}{2014}]{2014MNRAS.439.1628Z}
{Zu} Y.,  {Weinberg} D.~H.,  {Rozo} E.,  {Sheldon} E.~S.,  {Tinker} J.~L.,
  {Becker} M.~R.,  2014, \mn@doi [\mnras] {10.1093/mnras/stu033}, \href
  {http://adsabs.harvard.edu/abs/2014MNRAS.439.1628Z} {439, 1628}

\bibitem[\protect\citeauthoryear{{de Haan} et~al.,}{{de Haan}
  et~al.}{2016}]{2016ApJ...832...95D}
{de Haan} T.,  et~al., 2016, \mn@doi [\apj] {10.3847/0004-637X/832/1/95}, \href
  {http://adsabs.harvard.edu/abs/2016ApJ...832...95D} {832, 95}

\bibitem[\protect\citeauthoryear{{van Uitert} et~al.,}{{van Uitert}
  et~al.}{2017}]{2017MNRAS.467.4131V}
{van Uitert} E.,  et~al., 2017, \mn@doi [\mnras] {10.1093/mnras/stx344}, \href
  {http://adsabs.harvard.edu/abs/2017MNRAS.467.4131V} {467, 4131}

\makeatother
\end{thebibliography}



\appendix
\section{Morphological diversity of galaxy clusters in the current simulation}\label{sec:app}
\begin{figure*}
	\includegraphics[width=2\columnwidth]{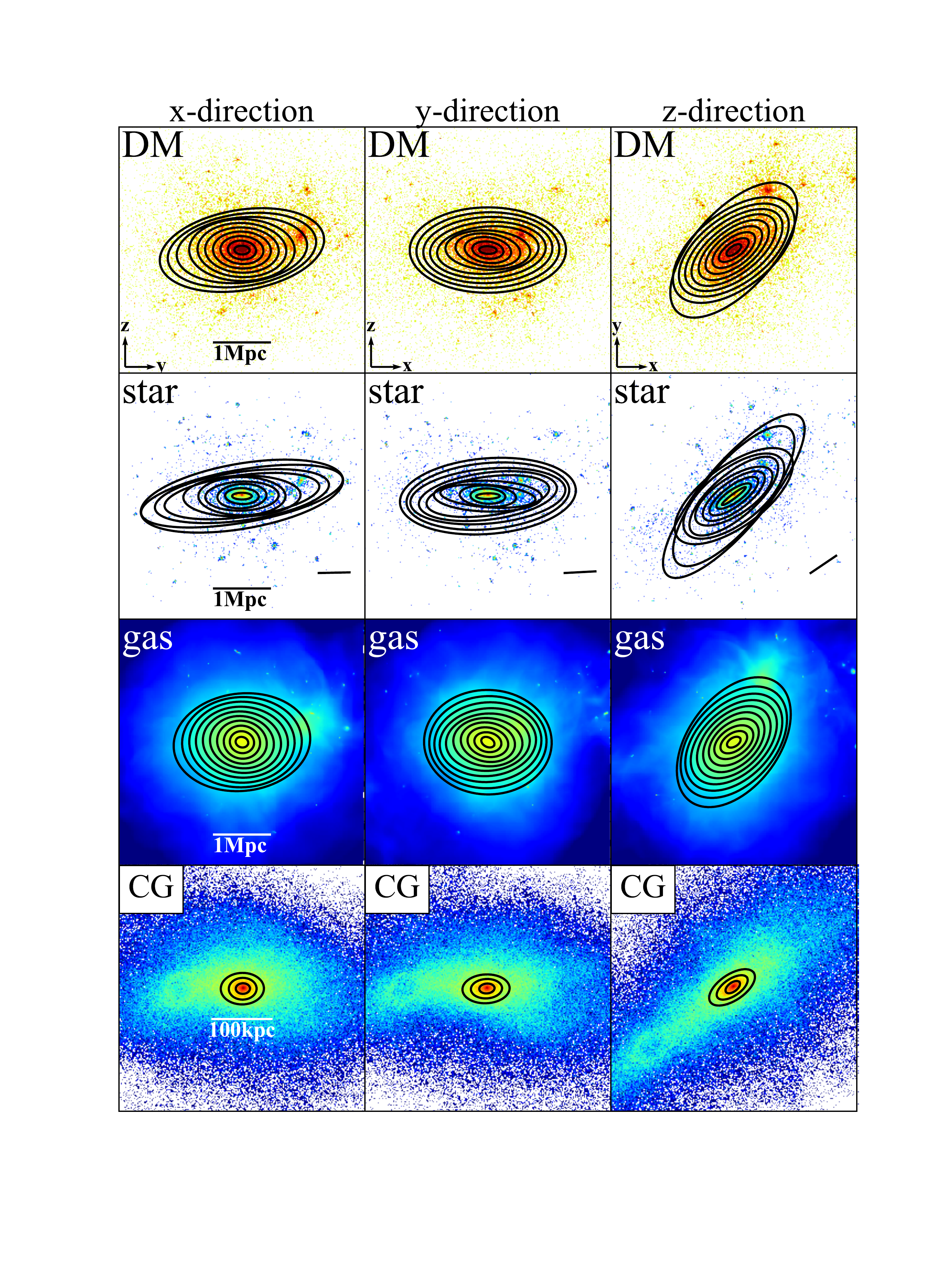}
    \caption{
    Projected distributions of dark matter, star, gas, and the CG from top to bottom of a galaxy cluster.
    This cluster is same as the one described in Section~\ref{sec:no2}.
    The sizes of the panels are $4.24$\,Mpc$\times4.24$\,Mpc for dark matter, star, and gas distributions and $400$\,kpc$\times400$\,kpc for the CG distributions.
    Left, centre, and right panels show images projected along $x$-, $y$-, and $z$-directions, respectively.
   Bars at right bottom in the star panels indicate the direction of the major-axis of the CG.
   }
    \label{fig:im2}
\end{figure*}

\begin{figure*}
	\includegraphics[width=2\columnwidth]{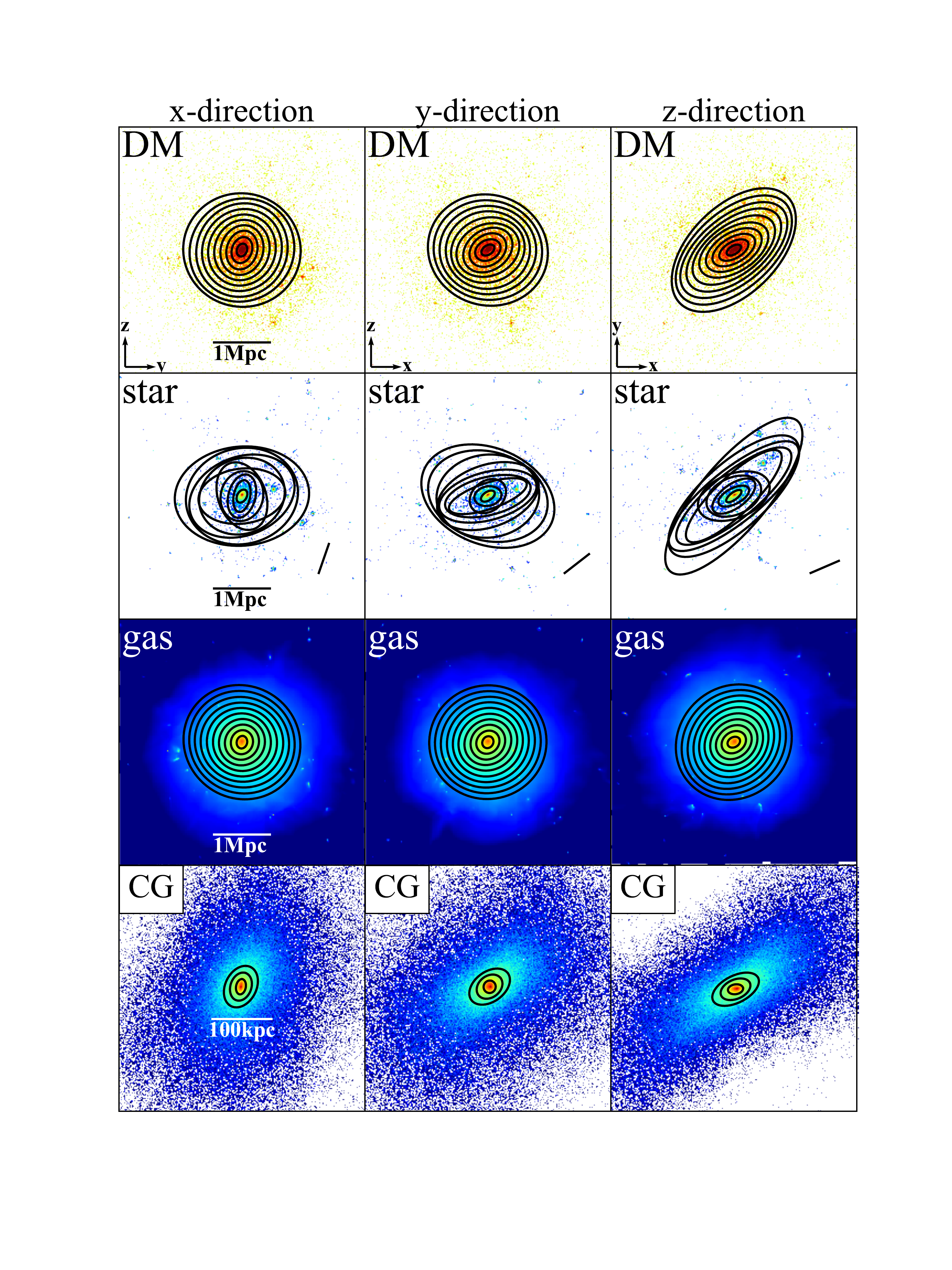}
    \caption{
    The same images as Fig.~\ref{fig:im2} but for an example of clusters having rounder shapes in projected images.
    }
    \label{fig:im3}
\end{figure*}

\begin{figure*}
	\includegraphics[width=2\columnwidth]{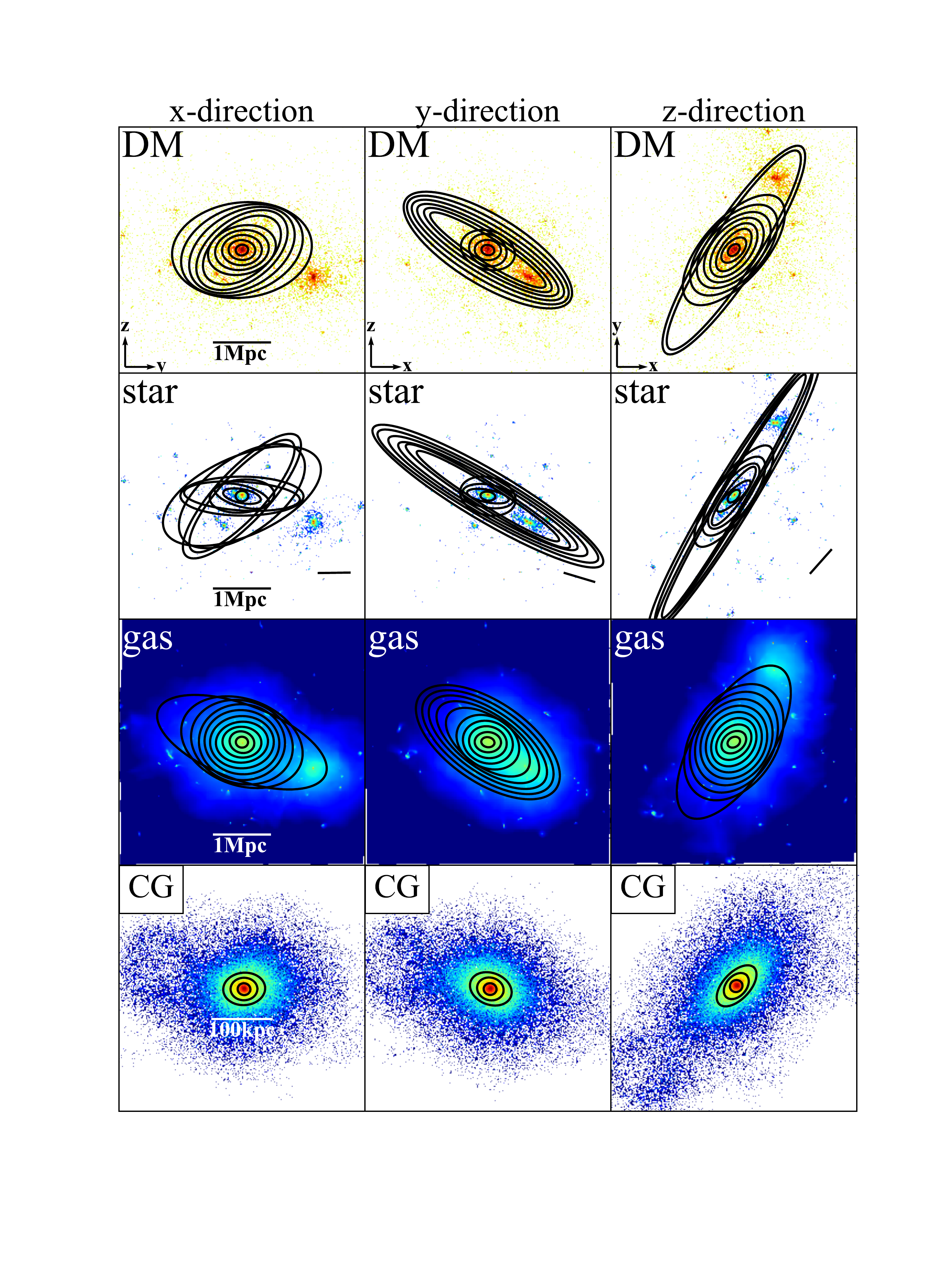}
    \caption{The same images as Fig.~\ref{fig:im2} but for an example of clusters having dominant substructures that significantly affect the ellipse fit.
    }
    \label{fig:im7}
\end{figure*}

We showed projected images for one galaxy cluster in Section~\ref{sec:no2} just as an example.
Here we show three representative galaxy clusters just to show morphological diversity of galaxy clusters that we analysed in this paper.
Fig.~\ref{fig:im2} shows projected images of DM, star, gas, and CG distributions of the cluster described in Section~\ref{sec:no2} 
but for three different projection directions.
The dark matter halo looks elliptical for all three line of sights, indicating that their three dimensional shape is triaxial.
Fig.~\ref{fig:im3} shows another example, which has relatively rounder shape of the dark matter halo.
Such a round cluster may be a relaxed cluster that have experienced the major-merger in the past.
In spite of such a round shape of dark matter halo, the star distribution is elliptical due to sub-structures. 
While dark matter distributions projected along $x$- and $y$-directions are circular, that projected toward $z$-direction is elliptical, 
clearly demonstrating that circular distributions in projected space do not necessarily 
indicate spherically symmetric distributions in three dimensional space.
Fig.~\ref{fig:im7} shows an example of clusters which have dominant substructures in dark matter distributions.
There is a dominant sub-halo in the dark matter distribution, which significantly distort the ellipses for all the projected images that are used to derive ellipticities and position angles.
We find that about one third of clusters we analysed have such dominant substructures.



\bsp	
\label{lastpage}
\end{document}